\documentclass[aps,prb,twocolumn,showpacs,superscriptaddress]{revtex4-2}
\usepackage{amssymb,amsmath}
\usepackage{graphicx}
\usepackage{multirow}
\usepackage{hyperref}
\usepackage{appendix}
\usepackage{mathtools}
\usepackage{amsmath}
\usepackage{xcolor}
\usepackage{siunitx}

\graphicspath{{./IMG/}}
\newcommand{\bs}[1]{\boldsymbol{#1}}
\newcommand{\ind}[1]{_\text{#1}}

\begin{document}

\title{Surface-plasmon polaritons in multilayer jellium systems: dispersion and spatial description}

\author{Alexandre Cloots}
\affiliation{European Theoretical Spectroscopy Facility, Institute of Condensed Matter and Nanosciences, Universit\'{e} catholique de Louvain, Chemin des \'{e}toiles 8, bte L07.03.01, B-1348 Louvain-la-Neuve, Belgium}
\author{Tanguy Colleu}
\affiliation{Laboratoire de Physique du Solide, Namur Institute of Structured Matter (NISM), University of Namur, 61, Rue de Bruxelles, B-5000 Namur, Belgium}
\author{Vincent Liégeois}
\affiliation{ELaboratory of Theoretical Chemistry, Unit of Theoretical and Structural Physical Chemistry, Namur Institute of Structured Matter (NISM), University of Namur, Rue de Bruxelles, 61, B-5000 Namur, Belgium}
\author{Gian-Marco Rignanese}
\affiliation{European Theoretical Spectroscopy Facility, Institute of Condensed Matter and Nanosciences, Universit\'{e} catholique de Louvain, Chemin des \'{e}toiles 8, bte L07.03.01, B-1348 Louvain-la-Neuve, Belgium}
\author{Luc Henrard}
\affiliation{Laboratoire de Physique du Solide, Namur Institute of Structured Matter (NISM), University of Namur, 61, Rue de Bruxelles, B-5000 Namur, Belgium}
\author{Xavier Gonze}
\affiliation{European Theoretical Spectroscopy Facility, Institute of Condensed Matter and Nanosciences, Universit\'{e} catholique de Louvain, Chemin des \'{e}toiles 8, bte L07.03.01, B-1348 Louvain-la-Neuve, Belgium}

\date{\today}

\begin{abstract}
Surface-plasmon polaritons (SPPs) are electromagnetic waves that propagate along metal-dielectric interfaces, with important applications in sensing, energy, and nanotechnology. While the behavior of SPPs in single metal slabs is well understood, the coupling between plasmon modes in multilayer systems has received less attention. 
In this paper, we explore the response functions of SPPs in single-slab, double-slab, and two-different-slab systems using the jellium model. 
Thanks to a comparison with classical models, our study reveals how quantum effects influence the resonance frequencies of these modes.
It also details the spatial description of the different SPP modes and unveils how their coupling occurs in two-different-slab systems. 
These findings provide new insights into the behavior of SPPs, especially in complex nanostructures.

\end{abstract}


\maketitle

\section{Introduction}

Plasmons, which arise from the resonant interaction between electromagnetic fields and electron gases~\cite{murray2007plasmonic,kravets2018plasmonic}, have garnered significant attention due to their impact on diverse technological applications. 
These include enhancing the sensitivity of biosensors~\cite{hill2015plasmonic}, improving the efficiency of photovoltaic cells~\cite{ferry2010design}, and enabling surface-enhanced vibrational spectroscopies (SEVS)~\cite{sharma2012sers, willets2007localized, stiles2008surface, Langer2019, Neubrech2017, colleu2024surface}. 
Plasmonic phenomena are also of particular interest in the field of photonics~\cite{barnes2003surface}, with recent advancements in graphene-based plasmonics offering new avenues for research~\cite{grigorenko2012graphene}.

Plasmons can be broadly categorized into two types: volume plasmons, which exist within a bulk material, and surface plasmons, which occur at the interface between a metal and a dielectric medium or vacuum~\cite{maier2007plasmonics}. 
Surface plasmons, in particular, exhibit unique properties that are especially advantageous for the aforementioned applications. When excited by an external electromagnetic field, surface plasmons result in a pronounced enhancement of the electric field near the material interface, leading to strong field confinement at the nanoscale~\cite{sharma2012sers}.

Two distinct modes of surface plasmons are typically distinguished: surface-plasmon polaritons (SPPs) and localized surface-plasmon resonances (LSPRs). Both involve the collective oscillation of free electrons within a metal. 
SPPs are propagating excitations that travel along the interface between a metal and a dielectric, whereas LSPRs are non-propagating excitations confined at the surface of metallic nanostructures significantly smaller than the wavelength of the incident light~\cite{maier2007plasmonics, willets2007localized}. 
These excitations generate a strong localized field enhancement in the vicinity of the nanostructure surface.
In this work, the focus is on the properties and behavior of SPPs within a jellium model and their implications for plasmonic applications.

The theoretical understanding of surface-plasmon phenomena was pioneered by Ritchie, who first described them using a macroscopic dielectric function approach~\cite{ritchie1957plasma}. 
Since then, significant advancements in both analytical and computational techniques have enabled a more comprehensive characterization of plasmon properties. Early analyses relied on classical electrodynamics, employing Maxwell's equations to describe the electromagnetic fields associated with plasmons~\cite{maier2007plasmonics}.
Subsequent developments included the hydrodynamic model, which approximates the electron gas as a fluid, and the jellium model, which provides a simplified treatment of the electron distribution in metals albeit retaining key quantum effects~\cite{eguiluz1975influence, eguiluz1980electron,eguiluz1981screening}.
More recently, first-principles calculations have been employed to capture the atomic-scale and quantum mechanical effects underlying plasmon behavior~\cite{andersen2012spatially}.

These theoretical frameworks, ranging from classical to quantum descriptions, have greatly expanded our understanding of plasmon properties~\cite{ferrell1957characteristic, powell1959origin, economou1969surface,newns1970dielectric, chen1992retarded, dobson1992electron}. A comprehensive overview of these developments up to the early 21st century can be found in the review by Pitarke \textit{et al.}~\cite{Pitarke:2006aa}.

The jellium model, which approximates a gas of free electrons moving in a uniform, positively charged background, has served as a cornerstone for understanding plasmonic behavior for several decades. This model allows one to focus on the collective electronic properties without the complexities of individual ion cores. While recent advances in computational power have enabled first-principles studies that can now simulate systems containing hundreds of atoms~\cite{Koval:2016aa, Titantah:2016aa, Kuisma:2015aa}—a significant leap compared to the limitations of a few atoms in the past century~\cite{manninen1986structures}—the jellium model remains a powerful and insightful tool for many scenarios. It continues to provide valuable insights into various surface and electronic properties, such as the work function~\cite{Lang:1970aa, Lang:1971aa}, surface-plasmon-resonance frequencies~\cite{ritchie1957plasma}, heterostructure modeling~\cite{asmar1992jellium}, and the behavior of metallic nanospheres~\cite{ekardt1984dynamical}, all at a significantly lower computational cost compared to first-principles approaches~\cite{pitarke2001jellium, liebsch1997electronic, dobson2004testing}.

Numerous studies have validated the accuracy and versatility of the jellium model, demonstrating that it provides reasonable results for a wide range of systems. 
Its limitations typically arise only when atomic lattice structures or band-specific effects become prominent. In such cases, more sophisticated adaptations, such as the stabilized jellium model~\cite{Pitarke:2006aa, liebsch1997electronic}, the incorporation of empirical atomic layer potentials~\cite{echarri2020theory}, or advanced screening models tailored for conduction band electrons in materials like silver~\cite{rocca1995surface}, offer substantial improvements without significantly increasing computational complexity. 
Thus, the jellium model, along with its modern refinements, is a well-established and  valuable framework for exploring and predicting the electronic properties of metals. 

Among the different approximations possible to describe the jellium electron gas interactions, the Random Phase Approximation (RPA), discarding the exchange-correlations effect in the response density, is most commonly used. This approximation has long been deemed a decent approach for the prediction of plasmonic phenomena in simple metals such as the alkaline ones\cite{liebsch1997electronic}.
%\textcolor{red}{XG20241011 : This sentence does not situate the RPA as being a flavor of the jellium model. This is confusing.}

The surface-plasmon polaritons, that are the focus of this paper, are observed experimentally and hence predicted theoretically as peaks in the loss spectra. Their signature can be derived from different approximations, adapted to various experimental situations. 
Typically, the theory decomposes the interaction of an electron or an electron beam impinging on a surface in excitations in the solid resulting from exchange of given wavevector and energies. As this electron travels parallel or through the surface, it excites the electron gas, leading to resonant behavior at certain frequencies. 
Note that plasmon modes can also be excited by any electromagnetic field, as long as they have the right frequency to trigger a resonant behavior.
%by photons or more generally
%Note that in the SPP case, the coupling with photons is trickier, it can nonetheless be achieved by either adding a prism as an intermediate layer between the dielectric and the metal or using a grating on the surface.

Surface-plasmon polariton dispersion relations have long received special attention because they give information about the frequency where a field enhancement can be expected at a given wavevector, which is essential for the proper functioning of applications utilizing plasmonics\cite{economou1969surface, ritchie1966surface, vincent1973dispersion, tsuei1991normal, penzar1984surface}. The knowledge of the conditions required to activate a plasmon is indeed a prerequisite to get, for example in SEVS experiment such as Electron Energy Loss Spectroscopy (EELS) or Attenuated Total Reflection, the signal enhancement at the frequencies of interest. Various methods have been developed to extract such dispersion relations from the response of the system, as it includes plain bulk effects as well. In this paper, two distinct approaches to predict the frequencies of surface-plasmon modes are considered. 

The first is based on the \textit{surface response function} (SRF). This function, derived by Persson \textit{et al.}\cite{persson1985electron}, allows one to predict all the resonance frequencies of surface modes and to obtain a spectrum that can be compared with EELS spectra. 
This method relies on the density response function. It is worth noting that several similar approaches, relying sometimes on other response functions, have been proposed in the literature to predict results corresponding to different experimental setups\cite{garcia2001energy, echarri2020theory}. 

A second approach based on the loss function associated with the macroscopic dielectric function, that will be called for the sake of brevity the \textit{macroscopic loss function} (MLF), is also used. The spectrum obtained  does not correspond to any experimental setup but to the EELS spectra of an homogeneous material described by the associated effective dielectric function. This common approach in quantum calculations based on supercell approach also delivers the decomposition into eigenmodes of the dielectric function, which allows for the easy identification of the change of electron density and potential associated with the eigenmodes (and thus characterization of their symmetry)\cite{andersen2012spatially}. Similar results can be obtained by using the definition of the density response function~\cite{echarri2020theory}.

In this paper, the jellium model is used as a platform to analyze the dispersion of SPPs and their coupling in systems of one or two slabs.
Single-slab systems have been studied in Refs.~\onlinecite{schulte1976theory, echarri2020theory, pitarke1998surface, silkin2004band, silkin2011low, schaich1994excitation} and first-principle investigations were also performed on single-slab systems in Refs.~\onlinecite{andersen2012spatially, giorgetti2020electron} while double-slab systems where the focus of the other publications~\cite{dobson2004testing, despoja2011nonlocal, despoja2006excitation}. 
%\textcolor{red}{XG20241011 : This sentence gives the impression that all these references were about the jellium model - for both single and double slabs. Is this true ?}
The case of a single slab is now well established in the textbooks using classical methods\cite{maier2007plasmonics}. First-principles calculations for particular slabs of noble metal~\cite{politano2008dispersion} or alkaline systems~\cite{andersen2012spatially} have also been performed and the results have been compared to the jellium model. 
At variance, the coupling between surface plasmons in two slabs has been less studied, especially at the quantum level. 
Results about the van der Waals forces between two slabs have been obtained with the jellium model~\cite{dobson2004testing, despoja2011nonlocal, despoja2006excitation} and the dispersion curves of double-slab systems have been predicted using a second-principles method, computationally more efficient than what is presented in this work but less flexible and yielding less insights about the collective modes\cite{despoja2006excitation}. 
The situation where several nanofeatures interact with each other occurs more often experimentally. It is therefore interesting to dig deeper in the understanding of the coupling of the different modes. 
%\textcolor{red}{XG20241011 : In the whole introduction, try to reduce the number of "we" and "our". Use the passive formulation, or find others formulations less personal...}
This work relies on a spatial description of the different SPP to unravel the couplings occurring in double-slab systems. 
It is important to mention that our numerical tool allows us to study a wide range of systems that can be described by a 1D potential such as single or multiple slabs.

The methodology is first applied to single slabs as a test case and to position the jellium approach in its RPA form with respect to classical and first-principles approaches. This simple case is also used to identify the strength and failures of the two approaches to predict the frequencies
of surface-plasmon modes (SRF and MLF). 

Then, with these insights, the analysis is extended to a double-slab scenario. 
The dispersion curves of coupled slabs are extracted and details about the shape and symmetries of each of the plasmon modes are given, providing in-depth knowledge of the excitation of these systems. 
Finally, systems consisting of two different slabs are considered and the conditions to observe a coupling of the plasmonics modes are discussed, which, to the best of our knowledge, had never been done before with such details. 
We find that the inclusion of quantum effects can lead to energy shifts of the SPP resonant frequency with respect to the classical approximation. 
%\textcolor{red}{XG 20241012 : By tuning the potential in the interlayer distance, the progressive suppression of tunneling effect is simulated. AND HERE WE WILL MENTION WHETHER THIS IS THE CAUSE OF THE DIFFERENCE IN FIG 5, BETWEEN CLASSICAL APPROXIMATION AND JELLIUM RESULTS.}

The paper is organized as follows. In Sec.~\ref{sec:theory}, the details of the jellium approach are described together with  the links with the response functions and the EELS spectra for both the SRF and MLF. 
Sec.~\ref{results} analyzes the case of a single slab, focusing on the main differences between the present jellium model and the classical or first-principles frameworks. 
Then, the coupling of surface-plasmon modes in a double-slab system is considered, with a full description of the dispersion curves for two identical slabs. 
Afterwards, the two-different-slab case is investigated. The conditions for the coupling of plasmon modes are identified with the description of how this coupling evolves along the dispersion line. Sec.~\ref{conclusion} summarizes our main results.

%%%%%%%%%%%%%%%%%%%%%%%%%%
\section{Theory}
\label{sec:theory}
%%%%%%%%%%%%%%%%%%%%%%%%%%

This section lays the theoretical foundation used to predict the dispersion curves presented in Sec.~\ref{results}. The choice of the potential is first debated. Then the different response functions are presented in the case of systems having a two-dimensional periodicity. Finally, the two approaches (SRF and MLF) to obtain the loss spectra highlighting the collective behavior occurring in such systems are described. 

The jellium is uniquely characterized by its electron density, $n\ind{e}$, or equivalently by the Wigner-Seitz radius \(r\ind{s}\)\cite{giuliani2008quantum}, which defines the average distance between electrons in the metal. The bulk plasma frequency is given by~\cite{giuliani2008quantum}:
\begin{equation}\label{eq:plasma_freq}
    \omega\ind{p}^2 = 4\pi n\ind{e}=\frac{3 }{r\ind{s}^3}.
\end{equation}
All the equations are given in atomic units.

\subsection{Potentials}

The wave functions of the 1D system, $\phi_l(z)$, and their associated energy levels, $\epsilon_l$, are obtained by diagonalizing the one-dimensional Hamiltonian:
\begin{equation}\label{Hamiltonian}
    \hat{H}\phi_l(z) = \Bigg[-\frac{\nabla^2}{2}+V\ind{tot}(z)\Bigg]\phi_l(z) = \epsilon_l\phi_l(z),
\end{equation}
where \( V\ind{tot}(z) \) is a potential representing the system. The non-self-consistent field (NSCF)\cite{Despoja:2005aa} and the self-consistent field (SCF)\cite{Lang:1970aa} cases are considered to build the potential. In the first case, the potential of the system is defined a priori, while in the second case, a positive background, $n^+(z)$, is drawn and the potential is deduced from the joint background profile and the electronic density. 
In both cases, regions with jellium and regions with vacuum are distinguished, all with $x-y$ translational invariance: the potential has only a $z$ dependence.

\begin{figure*}
\begin{tabular}{ccc}
(a) &  & \\
 & \multicolumn{2}{c}{\includegraphics[width=0.4\textwidth]{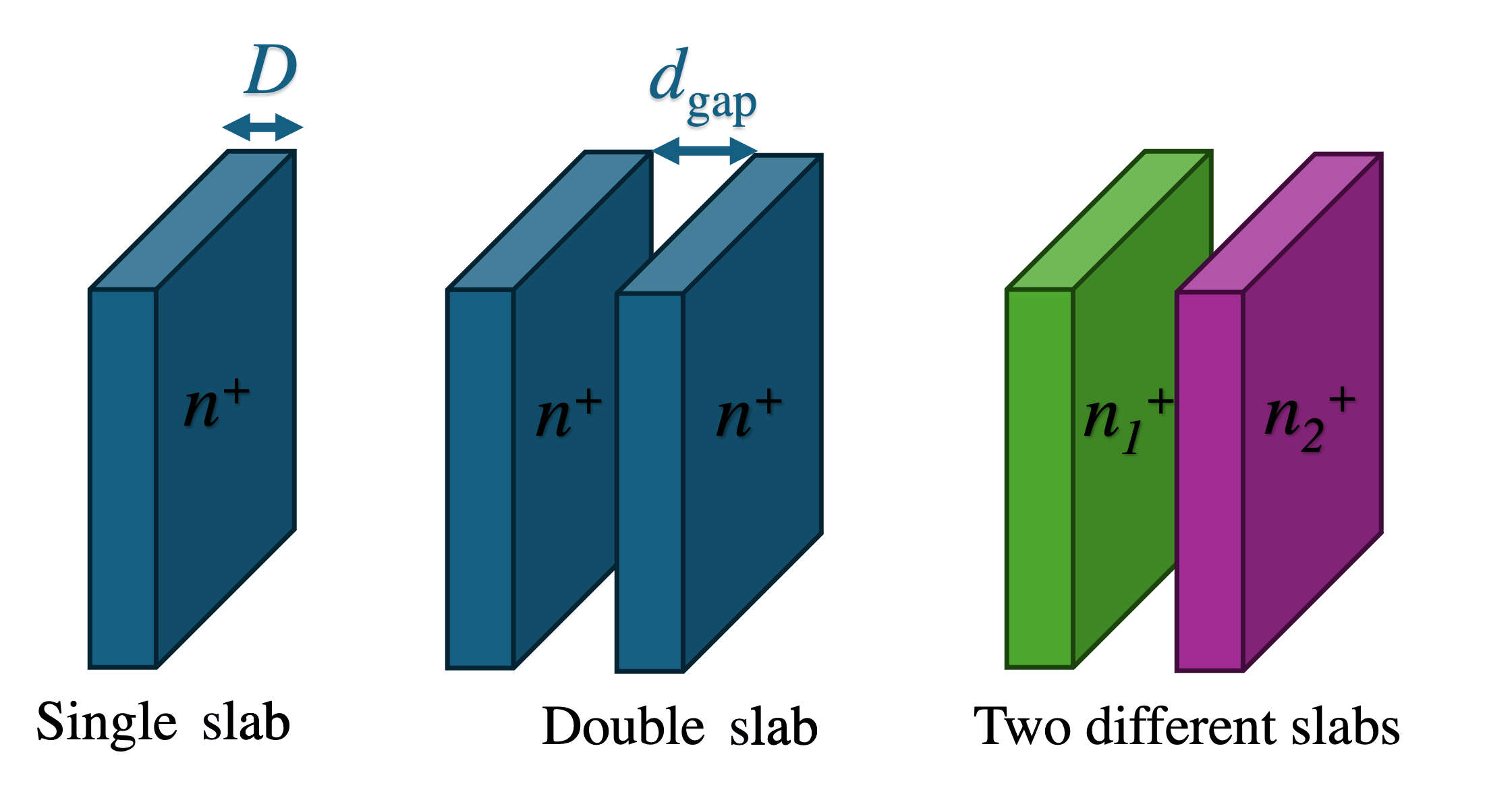} }   \\[12pt]
(b) &  &\\
 &  \includegraphics[width=0.4\textwidth]{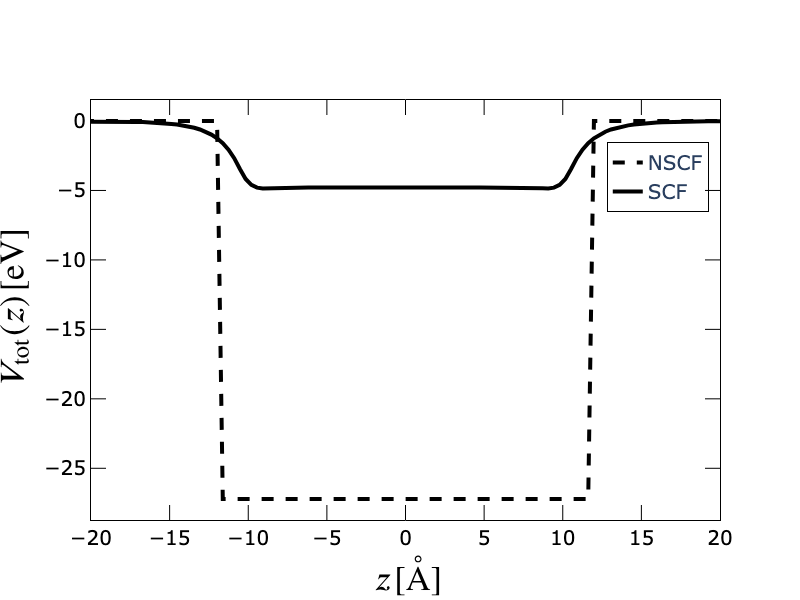} & \includegraphics[width=0.4\textwidth]{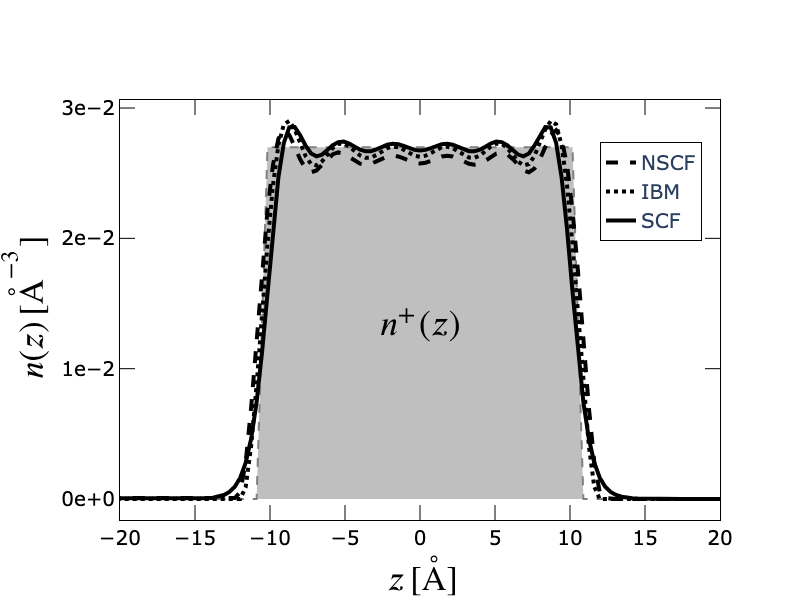}\\[6pt]
 (c) & & \\
 &    \includegraphics[width=0.4\textwidth]{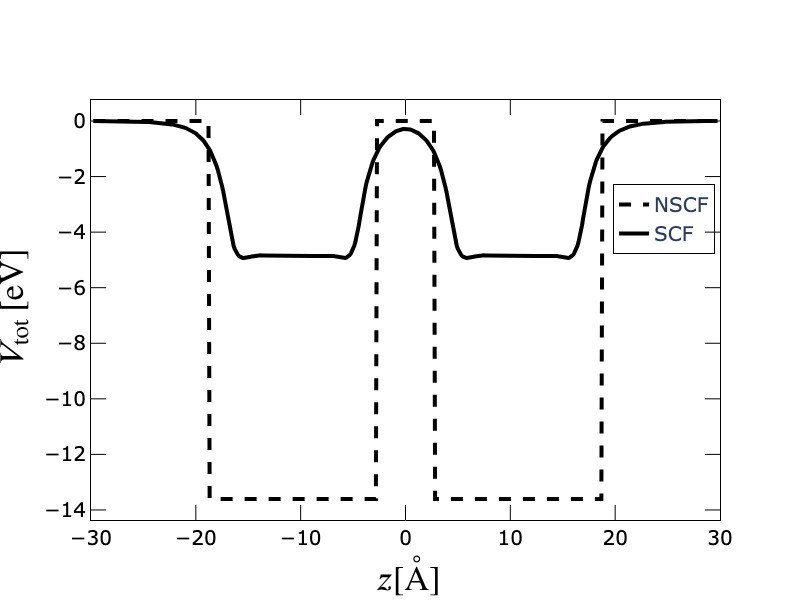} & \includegraphics[width=0.4\textwidth]{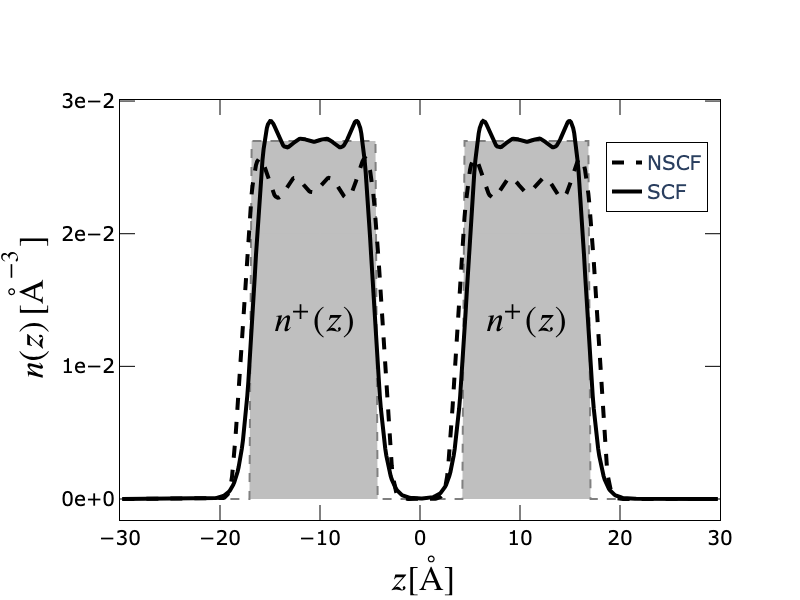}\\[6pt]
\end{tabular}
\caption{(a) The three systems studied in this paper: the single slab (left) of width $D$, the double slab with each slab showing the same width and background density $n^+$, and being separated by $d\ind{gap}$ (center), and the two different neutral slabs with the same width but with different background densities (right). (b) The different potentials (left) and their corresponding electronic densities (right) in the case of a single slab reproducing 10 atomic layers of Na. The width is $D = 21$~\si{\angstrom} ($n^+ = 0.027$~\si{\angstrom}$^{-3}$). (c) The different models for the potential (left) and their corresponding electronic densities (right) in the case of the double slab each reproducing 6 atomic layers of Na. The width is $D = 12.6$~\si{\angstrom} ($n^+ = 0.027$~\si{\angstrom}$^{-3}$) and $d\ind{gap}$=8.47~\si{\angstrom}. The dashed curves correspond to the NSCF scenario and the full curves are the result of the SCF approach.}
\label{fig:potential}
\end{figure*}

Moreover, in this paper, we restrict ourselves to the generic cases of one single slab and two coupled slabs, although the present approach can be adapted to more complex systems with 1D potential. 
%(many more could be considered) XG20240825 This it too obvious to be mentioned.
In these generic cases, shown in Fig.~\ref{fig:potential}a, one assumes that the transition region between the bulk ionic background and the vacuum is infinitely thin. 
The density of this ionic background is defined to reproduce a neutral material. 

Generally speaking, the total potential is composed of four terms: 
\begin{equation}
    V\ind{tot}(z) =  \Big(V(z)+ V\ind{Hxc}(z)\Big) + \Big(\delta V\ind{ext}(z) + \delta V\ind{Hxc}(z)\Big).
\end{equation}
The first one $V(z)$ is the potential that defines the system under study in the NSCF case.
In the SCF case, a second term $V\ind{Hxc}(z)$ is added, the Hartree and exchange-correlation potential.
It is the sum of the Hartree potential, $V\ind{H}(z)$:
\begin{equation}\label{eq:hartree}
    V\ind{H}(z) = -2\pi\int (n(z')-n^+(z'))|z-z'|dz',
\end{equation}
with $n(z)$ the electronic density profile, and the exchange-correlation potential $V\ind{xc}(z)$:
\begin{equation}
    V\ind{xc}(z)=\frac{\delta [e\ind{xc}(z).n(z)]}{\delta n(z)}.
\end{equation}
The Wigner LDA approximation is chosen for the exchange-correlation energy $e\ind{xc}(z)$:
%\textcolor{red}{XG20240825 Problematic ! $E\ind{xc}$ is a 1D integral in the present context of the  2D jellium. So, actually, should be formulated as a 3D integral over a repeated cell, divided by the surface area. Perhaps better, in the context of the present work, bypass $E\ind{xc}$.}.
\begin{equation}
    e\ind{xc}(z) = -\frac{0.458}{r\ind{s}(z)}+\frac{0.44}{r\ind{s}(z)+7.8},
\end{equation}
where
\begin{equation}
    r\ind{s}(z)=\Big(\frac{3}{4\pi n(z)}\Big)^{1/3}
\end{equation}
is the radius associated with the sphere occupied by an electron in the jellium.
In the NSCF case, $V\ind{Hxc}(z)$ vanishes. In the SCF case, $V(z)$ is used only as a first guess, and is then merged with $V\ind{Hxc}$ after the first iteration, the latter being the sole non-zero term in $V\ind{tot}$ at the end of the cycle.
The third term, $\delta V\ind{ext}(z)$ is an external perturbation, which produces a modification of the system.
$\delta V\ind{Hxc}(z)$ is the variation of the Hartree and exchange-correlation potentials (see later for the corresponding computation), induced by the variation of the density. 
When the system is perturbed by an applied potential, the response functions include the feedback from the electronic density variation (see Section~\ref{sec:resp_func}). 
%. Here, we restrict ourselves to one of the two potentials described below. 
%Adjusting the effective mass (not done here, the electron mass, $m_e=1$, gives another opportunity to adapt the hamiltonian to more complex systems). XG20240815 : not needed, this is just diverting the  discourse.

From the knowledge of the wave functions, and relying on the periodicity in the $x-y$ direction, the electronic density, $n(z)$,  can be calculated as follows\cite{Pitarke:2006aa}:
\begin{equation}\label{eq:density}
    n(z) = \frac{1}{\pi}\sum_i(\epsilon_F-\epsilon_i)|\phi_i(z)|^2\theta(\epsilon_F-\epsilon_i),
\end{equation}
%\textcolor{red}{XG20240825 Use the notation $n_{tot}$ to be consistent with the section B. Give a reference.}

with $\epsilon\ind{F}$ the Fermi energy. To ensure that the system is neutral, the following criterion is imposed:
\begin{equation}
    \int_{-L/2}^{L/2}[n^+(z)-n(z)]dz = 0,
\end{equation}
where $L$ is the width of the entire system (jellium and vacuum). The non-neutral case has been analyzed by Dobson~\cite{dobson1992electron}. Note that the value of the electronic density in Eq.~\ref{eq:plasma_freq} is equal to the average value of the electronic density in the slab, $n\ind{e}=1/D\int n(z)dz$.

In the case of the NSCF model, the potential is defined relative to the ionic background profile. For example, in the case of a single slab of width $D$ centered at $z=0$ (i.e. the background charge spans from $-D/2$ to $D/2$), the potential is defined as follows:
\begin{equation}\label{eq:NSCF}
    V(z) = \begin{cases}
        0 \quad &|z|>D/2+d\\
        C\quad & |z|< D/2+d,\\
    \end{cases}
\end{equation}
where $C$ is a negative constant that characterizes the jellium with respect to the vacuum. Its value will impact the 
spill-out of the electrons from the jellium to the vacuum. In a neutral system, $d=(3/16)\lambda_F$, with $\lambda_F$ the Fermi wavelength directly linked to the Wigner-Seitz radius, can be considered as a reasonable value to realistically account for the range of the spill-out of the electronic gas. 
In the case of a double-slab system, the single-slab potential is duplicated leaving a space $d\ind{gap}$ between the two wells. 

In the case of the SCF model, the potential $V(z)$ is chosen as an initial guess $V\ind{tot}^0(z)$. Then, Eq.~(\ref{eq:density}) is used to calculate the initial density $n^0(z)$. Since these two quantities do not satisfy the Poisson equation, a self-consistency cycle is started whereby, at the iteration $j+1$, the potential $V\ind{tot}^{j+1}(z)$ is calculated from the density $n^j(z)$ and the density $n^{j+1}(z)$ is obtained using the potential $V\ind{tot}^{j+1}(z)$ in Eq.(\ref{eq:density}), until they meet that requirement. To avoid any convergence issues in the SCF cycle, we rely on a simple mixing algorithm with a preconditioner in which the potential is updated as follows\cite{gonze1996towards}:
\begin{eqnarray}
\label{eq:scf}
    V\ind{tot}^{j+1}(z) &&=
    V\ind{tot}^j(z)+
    \nonumber\\
    &&\int \varepsilon^{-1}(z, z')\Big(V\ind{tot}[n^j](z')-V\ind{tot}^j(z')\Big)dz'.
    \nonumber\\
\end{eqnarray}
The preconditioner $\varepsilon^{-1}(z, z')$ is the fully periodic, static value, of the inverse dielectric function $\varepsilon^{-1}(z, z'; q = 0, \omega = 0)$ which is defined in the next subsection (Eq.~(\ref{eq:inv_diel})). 

Figure~\ref{fig:potential} illustrates the two potential models (NSCF and SCF) with their associated densities. Figure~\ref{fig:potential}b illustrates the single-slab scenario while Fig.~\ref{fig:potential}c shows the double-slab case. 

Another usual model found in the literature is the Infinite Barrier Model (IBM), in which the $C$ parameter of our NSCF approach is set equal to $-\infty$. The SCF and the IBM density profiles (the latter is also shown in Fig.~\ref{fig:potential}b for a single slab case, for information) are very similar but the SCF and IBM potentials are very different as the vacuum is not accessible for the electrons in the second case, making the excitations (electron-hole and plasmons) not correctly predicted. This justifies the choice of a NSCF potential rather than an IBM one. 
It allows one to tune the spill-out and the work function to better match an SCF system .

\subsection{Response functions}\label{sec:resp_func}

Different response functions can be defined to analyze the SPP excitations of surfaces and slabs.
Those functions describe the response of the system to an external perturbation. 
For example, the perturbation can be a time-dependent electric field, resulting in a time-dependent modification of the charge density. 
The irreducible (or \textit{non-interacting}) density response links the total variation of the potential at point $z'$  to the total variation of charge density at coordinate $z$.
It has been obtained by Eguiluz~\cite{eguiluz1985self} for Hartree electrons (so not taking into account exchange and correlation), for systems having a 2D periodicity\cite{1}:
\begin{multline}
    \chi^0(\bs{q}_{||}, \omega; z, z') =  \frac{\delta n\ind{tot}(\bs{q}_{||}, \omega; z)}{\delta V\ind{tot}(\bs{q}_{||}, \omega;z')}\\ =\sum_{l=1}^{l_M}\sum_{l'=1}^{\infty}F_{ll'}(\bs{q}_{||}, \omega)\phi_l(z)\phi_{l'}(z)\phi_{l}(z')\phi_{l'}(z'),
\end{multline}
with $l$ and $l'$ the different electronic subbands of the slab, and $\bs{q}_{||}$ the wavevector of the perturbation parallel to the surface, so in the $x-y$ plane. $l_M$ is the maximum occupied subband of the slab.  
%\textcolor{red}{XG20240825 This will become obvious with the notations are more consistent. So this sentence might be suppressed.}
\begin{multline}\label{Fll}
    F_{ll'}(\bs{q}_{||}, \omega) = -\frac{1}{ A}\sum_{\bs{k}_{||}}f_{\bs{k}_{||}l}\Bigg[\frac{1}{\bs{q}_{||}\cdot\bs{k}_{||}+a_{ll'}(\bs{q}_{||})+\omega+i\eta}\\+\frac{1}{\bs{q}_{||}\cdot\bs{k}_{||}+a_{ll'}(\bs{q}_{||})-\omega-i\eta}\Bigg],
\end{multline}
with $f_{\bs{k}_{||}l}$ the Fermi-Dirac filling of a given state for a given $\bs{k}_{||}$, a reciprocal vector parallel to the surface of the slab\cite{2}, $A$ the area of the surface of the slab, and
\begin{equation}
    a_{ll'}(q_{||}) = \frac{q_{||}^2}{2}-(\epsilon_l-\epsilon_{l'}),
\end{equation}
with $q_{||} = |\bs{q}_{||}|$. 
In the present case, since $\bs{q}_{||}$ is defined in the $x-y$ plane where there is a rotational invariance and since the sum over $k_{||}$ in Eq.~(\ref{Fll}) can be replaced by a factor $2A(E_F-\epsilon_l)/(2\pi)$, the response function can be expressed as a function of the norm of the wavevector.
%\textcolor{red}{XG20240825 : This is the first appearance of $q_{||}$, should be defined. Moreover, should mention that there is a rotational invariance, so that the vectorial character of $\bs{q}_{||}$ is irrelevant, and will not be emphasized anymore in what follows.}

The irreducible density response function being computed, one can obtain the dielectric function, $\varepsilon$, as well as its inverse (already used as a preconditioner in Eq.~(\ref{eq:scf})). Both quantities, $\varepsilon$ and $\varepsilon^{-1}$,  link the external perturbation $\delta V\ind{ext}$ to the variation of the total potential 
$\delta V\ind{tot}$ in the system. The subscript ``ext'' indicates that it is purely the applied variation while the ``tot'' one also takes into account the reaction from the system. It is the total difference of the potential in the system after and before the perturbation is applied. 
%\textcolor{red}{Should define again a Schrödinger equation, because in the NSCF, I think that there IS a H and XC component to the response ! This might be quite confusing ... To be discussed ...}\textcolor{blue}{AC20240830 : is it clearer like?}

\begin{widetext}
The dielectric function is defined as:
\begin{equation}\label{eq:epsilon}
    \varepsilon(z, z';q_{||}, \omega) = \frac{\delta V\ind{ext}(q_{||}, \omega;z)}{\delta V\ind{tot}(q_{||}, \omega; z')}= \delta(z-z')-\int dz''  
    \Big(v\ind{c}(z, z''; q_{||})+f\ind{xc}[n_0](z, z'';q_{||},\omega)\Big)
    \chi^0(z'', z'; q_{||}, \omega),
\end{equation}
and its inverse as:
\begin{equation}\label{eq:inv_diel}
    \varepsilon^{-1}(z, z';q_{||}, \omega) =\frac{\delta V\ind{tot}(q_{||}, \omega;z)}{\delta V\ind{ext}(q_{||}, \omega;z')}= \delta(z-z')+\int dz''
    \Big(v\ind{c}(z, z''; q_{||})+f\ind{xc}[n_0](z, z'';q_{||},\omega)\Big)
    \chi(z'', z'; q_{||}, \omega),
\end{equation}
with $v\ind{c}(z, z';q_{||})$ the Coulomb potential, and
$f\ind{xc}[n_0]=\frac{\delta V\ind{xc}[n]}{\delta n}\Big|_{n=n_0}$,
the exchange-correlation kernel, where $n_0$ is the unperturbed density.

In the particular geometry considered here, with planar translational invariance, the Coulomb potential is obtained as the inverse Fourier transform on the $q_z$ component of the 3D potential:
\begin{equation}
    v\ind{c}(z, z';q_{||}) = \frac{2\pi}{q_{||}}e^{-q_{||}|z-z'|}.
\end{equation}
It reduces to the $q_{||}=0$ limit, $v\ind{c} = |z-z'|$ if the charge neutrality is taken into account, as in Eq.~(\ref{eq:hartree}).

The reducible density response function $\chi$ is obtained by means of a Dyson equation, 
\begin{multline}
    \chi(z, z'; q_{||}, \omega) = \frac{\delta n\ind{tot}(q_{||}, \omega;z)}{\delta V\ind{ext}(q_{||}, \omega;z')}=\\\chi^0(z, z'; q_{||}, \omega)+\int dz_1\int dz_2\chi^0(z, z_1; q_{||}, \omega)\Big(v\ind{c}(z_1, z_2, q_{||})+f\ind{xc}[n_0](z_1, z_2;q_{||},\omega)\Big)\chi(z_2, z'; q_{||}, \omega).
\end{multline}
It will be denoted as \textit{exact} if both the coulomb and the exchange-correlation are integrated in the equation. If one applies the Random-Phase Approximation (RPA) in which the  exchange-correlation kernel vanishes $f\ind{xc}[n_0]=0$, the resulting function is called the \textit{interacting} density response function.%\textcolor{red}{XG20241012: such denomination seem incorrect to me. No need to postulate the RPA to have an interacting density response function. Where does this come from ? }. 

Finally, the screened interaction $W$ gives the change of 
total potential at $z$ induced by an external charge at $z'$,
\begin{equation}\label{eq:scr_int}
    W(z, z'; q_{||}, \omega) =\frac{\delta V\ind{tot}(q_{||}, \omega;z)}{\delta n\ind{ext}(q_{||}, \omega;z')}= v\ind{c}(z, z'; q_{||})+\int dz_1\int dz_2 v\ind{c}(z, z_1; q_{||})\chi(z_1, z_2; q_{||}, \omega)v\ind{c}(z_2, z';q_{||}).
\end{equation}
\end{widetext}

This set of equations was implemented numerically in the matrix formalism (where the rows and columns of the matrix correspond to the $z$ coordinates). In the RPA ($f\ind{xc}=0$), the Eq.~(\ref{eq:epsilon}) for a given ($q_{||}, \omega$) becomes:
\begin{equation}\label{eq:mat_eps}
    \underline{\bs{\varepsilon}} = \underline{\bs{I}} - \underline{\bs{v}}\ind{c}\underline{\bs{\chi^0}}.
\end{equation}
The inverse dielectric function is obtained by inverting this matrix. A numerically stable way to compute the reducible density response function is to extract it from
the right-hand side of Eq.~(\ref{eq:inv_diel}):
\begin{equation}
    \underline{\bs{\chi}} = \underline{\bs{v}}\ind{c}^{-1}(\underline{\bs{\varepsilon}}^{-1}-\underline{\bs{I}}).
\end{equation}

Finally, the screened interaction from Eq.~(\ref{eq:scr_int}) is computed as:
\begin{equation}\label{eq:mat_W}
    \underline{\bs{W}} = \underline{\bs{v}}\ind{c} - \underline{\bs{v}}\ind{c}\underline{\bs{\chi}}\underline{\bs{v}}\ind{c}.
\end{equation}

In the case where the wavefunctions and energies are computed based on the SCF potential, this algorithm to obtain the response functions is called \textit{SCF-based RPA}. 

\subsection{Surface response function (SRF)} 

In this section, we present a first approach to obtain loss spectra. 
It is more closely related to an experimental setup than the one using the macroscopic dielectric function, presented in Sec.~\ref{sec:mlf}. 
The justification of the use of the two methods is that the first method leads to a good understanding of the system based on actual experimental setups and focusing on the surface mode. 
The second method allows one to characterize better the modes found in the first method, irrespectively of the experimental setup (see Sec.~\ref{sec:mlf}). 

The SRF allows one to extract a spectrum similar to the one obtained in an EELS experiment. It evaluates the loss of energy of an electron traveling close to a surface, losing quanta of energy, mainly by exciting the associated plasmon.  The SRF ($g(q_{||}, \omega)$) can be computed from the interacting density response functions defined in the previous section. 
%This function, which plays a similar role for surface excitations as the longitudinal dielectric function for bulk excitations (i.e. its imaginary part presents peaks when a collective excitation occurs), 
It is given by~\cite{Pitarke:2006aa}:
\begin{equation}\label{eq:srf}
    g(q_{||}, \omega) = -\frac{2\pi}{q_{||}}\int dz_1 \int dz_2 e^{q_{||}(z_1+z_2)}\chi(z_1, z_2;q_{||}, \omega).
\end{equation}

%\textcolor{red}{XG20240825 : please, improve the style of this sentence and clarify it ...} The first of the two methods used in this paper to analyze the collective excitations of 1D profile systems is composed of two parts: (1) the eSRF gives a spectra where the position of the different excitations can be seen and (2) the imaginary part of the screened interaction, key function in the reproduction of the EELS experiment spectra. It is also used to obtain spatial information and categorize the excitation in bulk and surface plasmons (without dealing with the symmetry at this stage). 

The screened interaction of Eq.~(\ref{eq:scr_int}) for coordinates $z, z'>0$ and far from the surface (so, where the electronic density is null), 
%\textcolor{red}{XG20241012 : One needs to be more explicit about the values of $z$ considered to be "far from the surface". Should they be largely negative, or largely positive ? Note that the sign of $z1$, $z2$ or $z$ and $z'$ is important in both Eqs.(19) and (20)...} 
can, in this case, be expressed as a function of this SRF\cite{garcia2001energy}:
\begin{equation}\label{eq:g_to_W}
    W(z, z'; q_{||}, \omega) = v\ind{c}(z, z', q_{||})-\frac{2\pi}{q_{||}}e^{-q_{||}(z+z')}g(q_{||}, \omega).
\end{equation}
The rate at which the external potential generates electronic excitation in the many-body system is defined from Fermi's golden rule:

\begin{equation}
    w(\omega) = 2\pi\sum_{\bs{k}, \bs{k'}}f_{\bs{k}}(1-f_{\bs{k'}})|\langle \bs{k'}|V\ind{tot}|\bs{k}\rangle|^2\delta(\epsilon_{\bs{k}}-\epsilon_{\bs{k'}}-\omega).
\end{equation}
This rate can be related to the response functions thanks to~\cite{liebsch1997electronic}:
\begin{equation}
    w(\omega) = -2 \, \text{Im}\int d^3r V\ind{ext}^*(\bs{r}, \omega)\delta n(\bs{r}, \omega).
\end{equation}

Using the response functions above and with an external potential of the shape\cite{liebsch1997electronic}:
\begin{equation}
    V\ind{ext}(\bs{q_{||}}, \bs{r}) = -\frac{2\pi}{q_{||}}\exp(i\bs{q_{||}}\cdot \bs{r_{||}})\exp(q_{||}z),
\end{equation}

that describes the field generated by a charged particle traveling far from the surface in the $z>0$ region, it reduces to:
\begin{equation}\label{eq:wqw}
    w(q_{||}, \omega) = \frac{4\pi}{q_{||}A}\text{Im}\,g(q_{||}, \omega).
\end{equation}

Hence, the peaks in the imaginary part of the SRFs are associated with both interband (one electron) and intraband (collective)  excitations. This function can be compared with the results of an EELS experiment as the resonances in both cases are associated to the electron energy loss.  

The EELS spectra can also be computed for perpendicular or parallel trajectories of the electrons from the screened interaction response function or the reducible density response function, as detailed in Echarri \textit{et al.}\cite{echarri2020theory} or Garcia-Lekue \textit{et al.}\cite{garcia2001energy}. 
Here, we limit ourselves to the study of the imaginary part of the screened interaction for $z=z^\prime$ and the SRF as they are the leading term to predict EELS spectra and we are not comparing our results to a specific experimental result. 
The aim is rather to gather spatial information about the plasmon mode. 
%If the energy is scattered close to the surface or at the core of the system, it is, in our case, respectively, the result of the interaction with a SPP or a bulk plasmon. 
Note that for the simulation of an EELS spectrum, the response is analyzed at the electron position ($z=z^\prime$) and that this position can be inside the slab as well as outside the slab. 
In the latter case, only SPP is excited while in the former case volume plasmon dominate. The analysis of $\text{Im}\,W(z_0, z; q_{||}, \omega)$ for a $z_0$ close to the surface\cite{3} allows one to highlight the symmetries of the surface mode, which is not possible if only the $z=z^\prime$ response function is known. 
The use of the screened interaction function to obtain the spatial description of the modes is then less straightforward (it does not show all the modes and symmetries at once) than the one proposed in Sec.~\ref{sec:mlf}.

\subsection{Macroscopic loss function (MLF)}
\label{sec:mlf}

The second approach used in this paper is the computation of the MLF. As stated earlier, and contrary to the previous section, this function does not correspond directly to any experimental geometry. 
%A spectra is shown only to highlight the differences with the one obtained from the approach using the SRF. 
Nonetheless, this approach is interesting to acquire a better understanding of the different plasmon modes involved
as the spatial density change can be described easily. 
The macroscopic dielectric function (spatial average of the dielectric function, hence taking into account the vacuum in the cell in our case) is the basic quantity to extract the relevant information. Indeed, it is well known that plasmon modes occur at the poles of $\varepsilon\ind{M}^{-1}(q, \omega)$, i.e. where the real part of the dielectric function is small. It is also the quantity that is commonly extracted from ab-initio simulations of a slab, in a supercell geometry.

To obtain a spatial representation of each mode, Andersen~\textit{et al.}
\cite{andersen2012spatially}
proposed an algorithm that uses the fact that collective oscillations are associated with poles of the inverse dielectric function eigenvalues, $\varepsilon^{-1}_i$, i.e., where $\text{Im}\,\varepsilon^{-1}_i(q, \omega)$ is a local maximum, while the associated eigenvectors are the variations of potential, $V$ and charge density, $\rho$ of the plasmon mode. These vectors are connected by the Poisson equation. The spectral representation of the dielectric function then writes, in mixed space $(q_{||},z)$\cite{andersen2012spatially}:
\begin{equation}
    \varepsilon(z, z'; q_{||}, \omega) = \sum_i\varepsilon_i(q_{||}, \omega)V_{i}(z ; q_{||}, \omega)\rho^*_{i}(z';q_{||}, \omega), 
\end{equation}

This representation of the function is employed as such visualization offers valuable insight on the evolution of the modes. 
%using coordinates of the direct and recirpocal space

%\textcolor{red}{XG20240825 : Since $z$ and $z'$ are arguments of $\varepsilon$, it would be better that they are arguments of the V and $\rho$, and not subsccripts.} 
It can also, alternatively, be expressed in reciprocal space.
Indeed, since the supercell formalism is being used to perform the computations (the supercell is made of jellium and vacuum), it gives equivalently\cite{4}:
\begin{equation}
    \varepsilon_{G, G'}(q_{||}, \omega) = \sum_i\varepsilon_i(q_{||}, \omega)V_{i, G}(q_{||}, \omega)\rho^*_{i, G'}(q_{||}, \omega),
\end{equation}
with $G$ and $G'$, the norm of the one-dimensional reciprocal vectors along the $z$ direction.

The MLF, denoted $L\ind{M}$, gives the loss incurred by electrons traveling in an hypothetical homogeneous medium described by $\epsilon_M$. The difference with the approach of the previous section is that the geometry of the system is not correctly described. It is therefore not possible to connect the spectrum to an experimental setup. The MLF is defined as:
\begin{equation}
    L\ind{M}(q_{||}, \omega) = -\text{Im}\,\frac{1}{\varepsilon\ind{M}(q_{||}, \omega)},
\end{equation}

The macroscopic dielectric function, $\varepsilon\ind{M}$, is the spatial average of the microscopic dielectric function (it is taken as the inverse of the constant component of the inverse dielectric function to include the local effects):
\begin{equation}
    \varepsilon\ind{M} = \frac{1}{\varepsilon^{-1}_{G=0,G'=0}(q_{||}, \omega)}.
\end{equation}
It can be decomposed in a weighted sum of its eigenmodes:
\begin{align}
    L\ind{M}(q_{||}, \omega) =& -\text{Im}\,\sum_iw_i\frac{1}{\varepsilon_i(q_{||}, \omega)}\\=& \sum_iw_i(q_{||}, \omega)L_i(q_{||}, \omega),
\end{align}
with 
\begin{equation}
    w_i(q, \omega) = V_{i, G=0}(q, \omega)\rho^*_{i, G=0}(q, \omega),
\end{equation}
and
\begin{equation}
    L_i(q_{||}, \omega) = -\text{Im}\,\frac{1}{\varepsilon_i(q_{||},\omega)}.
\end{equation}

Since the macroscopic quantities presented above are spatial average of their microscopic counterpart, the response functions should be renormalized to compare results obtained with different supercell sizes\cite{majerus2023anisotropy}
\cite{5}.

This algorithm then gives a global view of the plasmonic response and, 
thanks to its decomposition in eigenmodes, 
allows one to visualize the spatial variation of each mode through the eigenvector $V_i(z, q_{||})$ and $\rho_i(z, q_{||})$.
%\textcolor{red}{XG20240825 :  notations ? label them correctly.}

It can already be noted that the surface response and the MLFs spectrum do not present the same peak intensities.
Indeed, the weights used in the construction of the macroscopic loss are zero in the case of antisymmetric modes.
However, those can still produce a field enhancement in the vicinity of the surface and are interesting to characterize.
On the other hand, the decomposition in eigenmodes used in the build-up of the loss function allows one to retrieve the spatial description of all the modes (symmetric and antisymmetric alike).
This information is complementary and alleviates the need to use paired and unpaired wavefunctions in two sets as in Ref.~\onlinecite{penzar1984surface}.

\subsection{Numerical parameters}

The results in the following section were obtained using a grid of points along the $z$ direction with a spacing ranging from 0.18 to 0.5~\si{\angstrom} and with 40 to 120~\si{\angstrom} of vacuum, depending on the system under study (the larger the parallel wavelength, the larger the vacuum; the smaller the wavelength, the smaller the grid spacing). 
The damping parameter, $\eta$ in Eq.~(\ref{Fll}),
was fixed to 0.05~eV. 
Convergence with respect to the number of unoccupied bands was performed to guarantee an accuracy of 0.02 eV on the position of any plasmon peak position. The background density corresponds to the electronic density of Na, $n^+  = 0.004$~Bohr$^{-3}= 0.027$~\si{\angstrom}$^{-3}$.
The height of the barrier in the NSCF was set to 0.5 Ha $=$ 13.61~eV. It was chosen to reproduce as well as possible the plasmonic modes found with the SCF potential in the single slab case. A smaller height, closer to the experimental value of sum of the Fermi level and the work function, led to a red-shift of the mode while a larger value, closer to what one would use in the IBM case, led to a blue-shift of the surface peak, as the energy levels of the vacuum are not accessible. Throughout this paper, the self-consistent potential and density profiles were obtained after 5 to 10 cycles (convergence was considered to be achieved when $\frac{1}{L}\int |n_{i+1}(z)-n_i(z)|dz<10^{-7}$~Bohr$^{-3}\approx 6.74810^{-7}$~\si{\angstrom}$^{-3}$).

%%%%%%%%%%%%%%%%%%%%%%%%%%
\section{Results and Discussion}
\label{results}
%%%%%%%%%%%%%%%%%%%%%%%%%%

\subsection{Single slab}

This section explores the dielectric properties and collective excitations of a simple jellium slab model. 
The two approaches presented above to derive the electronic jellium profile are considered, the NSCF potential and the SCF potential. 
The differences between these approaches and their impact on the resulting dispersion relations are analyzed. The reference system aims at reproducing a 10 layer Na slab and is shown in Fig.~\ref{fig:potential}b. We use the first-principles value of 21~\si{\angstrom}\cite{andersen2012spatially} for $D$, the width of the slab. 
%\textcolor{red}{XG20240829 : how do you characterize that it is the jellium equivalent of a 10 layer Na slab ? - in terms of electronic density, and width, I guess. Could specify ? Why not also in terms of work function ?}\textcolor{blue}{If I use the work function value, the position of the surface plasmon peaks is worse, the density profile is more (too?) spread... The IBM gives a blue shift of the surface mode, the work function a red shift... the intermediate solution gives the right surface modes}

%\textcolor{red}{XG20240829 : here, we should understandlearn that Fig.1 systems is actually the jellium equivalent of a 10 layer Na slab. Mention it earlier, perhaps in the caption of Fig.~1}
As illustrated in the left part of Fig.~\ref{fig:potential}b, 
the ground-state electron density profiles for the two scenarios exhibit similar behavior within the slab, showing Friedel oscillations\cite{friedel1952xiv, schaich1994excitation} in both cases, but differ outside the surface with a more pronounced spill-out of the electrons in the NSCF case. 
The SCF potential, on the other hand, is very different from the NSCF one as shown on the left part of Fig.~\ref{fig:potential}b. The height of the barrier in the SCF case ($= 4.86$~eV) is closer to the sum of the real work function (experimental value $\approx 2.25$~eV\cite{wigner1997theory}, computed 1.49~eV) and real Fermi level (experimental value $\approx 3.24$~eV\cite{burchamashcroft}, computed 3.36~eV) than in the NSCF case, where the height was set to 13.61~eV. As mentioned above, this value was chosen to get the right position of the surface-plasmon peaks. 
Using a more realistic value for the work function in NSCF leads to a red shift of the peak positions. 
Additionally, some ripples mimicking the Friedel oscillations of the density can be seen at the bottom of the SCF potential which, by construction, are not present in the NSCF case.
%\textcolor{red}{XG20240829 : why was it set to 0.5 Ha? Shouldn't it be set to a value closer than the actual work function of the Na slab?}

\begin{figure}
\includegraphics[width = 0.5\textwidth]{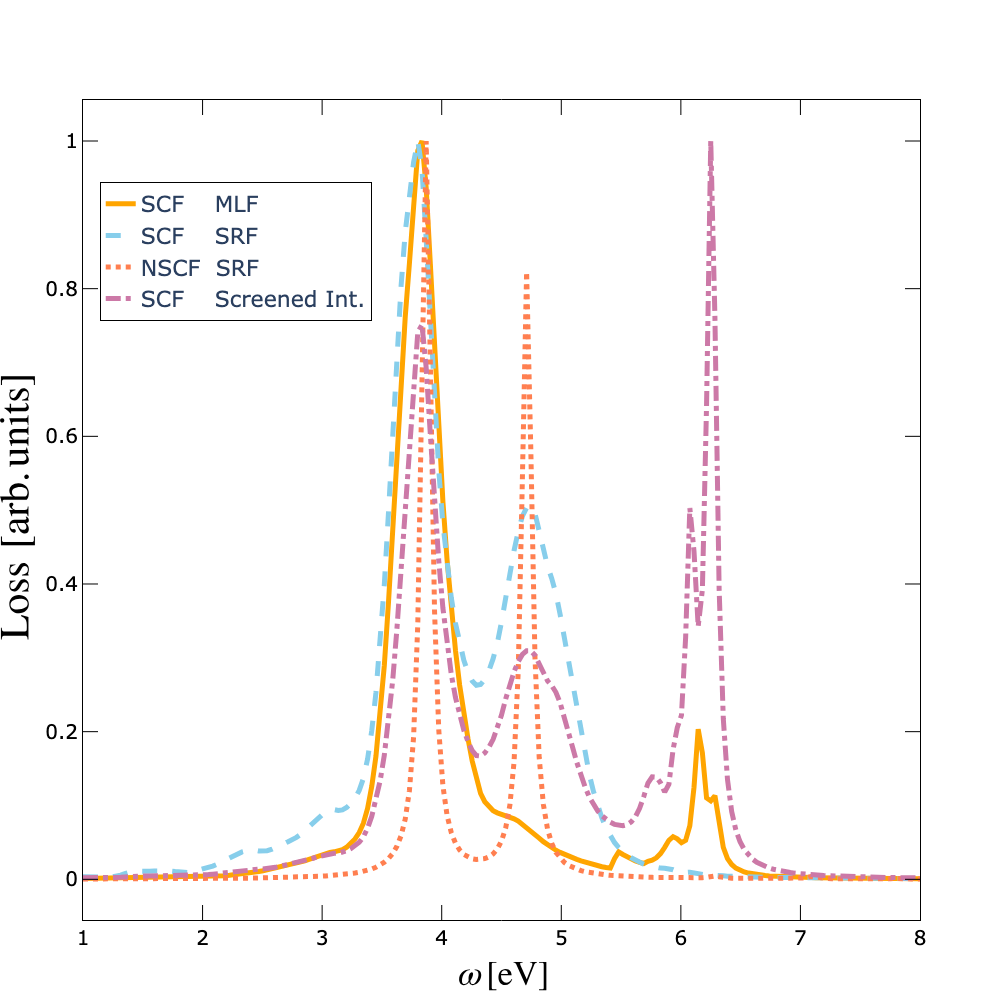}
\caption{Four spectra of energy losses. Three are from the two methodologies (MLF, SRF) with a comparison between the NSCF and SCF potential and one is obtained from screened interaction function. 
The spectra are shown for $q_{||} = 0.076$~\si{\angstrom}$^{-1}$. The spectrum for the NSCF case extracted from the SRF is in dotted coral, with two surface modes at 3.87 and 4.71~eV and one bulk mode (much less intense) at 6.29~eV. 
The spectra found with the SCF approach are in full line orange in the case of the MLF, showing one surface mode  at 3.83~eV and the bulk mode at 6.16~eV, and in dashed light blue in case of the SRF, showing two surface modes (3.80~eV and 4.71~eV).
In the latter case, the bulk mode is not seen on the figure. 
The spectrum obtained as the sum over all $z$ of the screened interaction, $\sum_{z}\text{Im}\,W(z, z; \omega)$ is in dot-dashed purple.
%, with three modes, much broader than in the other spectra. 
}
\label{fig:spectra}
\end{figure}

Figure~\ref{fig:spectra} presents typical spectra obtained from the various methods presented earlier, for $q_{||} = 0.076$~\si{\angstrom}: with SCF approach or not, SRF, MLF or with the screened interaction function (performing a sum over all $z=z'$ positions at each frequency). In those spectra, three main peaks can be identified. The two peaks with the lowest energy, around 3.83 and 4.71~eV are attributed to SPP while the third one at 6.16~eV (only visible with the MLF and the screened interaction methods) is a bulk plasmon. 
This identification is substantiated hereafter. 
Note that the spectra have been normalized to the maximum of the functions. 
A significant difference between the MLF and SRF spectra is the near disappearance of the second SPP peak in the former (only a shoulder in the first peak is visible). 
The second SPP peak in the SRF spectra can therefore be attributed to an asymmetric mode as explained in Sec.~\ref{sec:theory} and as corroborated by the visualization of the mode hereafter. 
The range of frequency close to the plasma frequency highlights also typical differences between the spectra. 
In the SRF, the bulk plasmon appears very weak as this function monitors the induced fields outside the slab. 
These are obviously a lot larger when an SPP is generated than when a bulk plasmon is excited. 
In the MLF and screened interaction spectra, the bulk plasmon peak appears around 6.16~eV. It is subdivided into two bulk modes (6.07~eV and 6.25~eV) which are preceded by a weak 'subsurface mode', around 5.8~eV. 
After checking that the screened interaction yields the same peak positions as the other methodologies, we infer that all the information required to perform the rest of our analysis are contained in the SRF and MLF methodology.

\begin{figure*}
\begin{tabular}{cccc}
(a) & & (b) & \\
& \includegraphics[width = 0.45\textwidth]{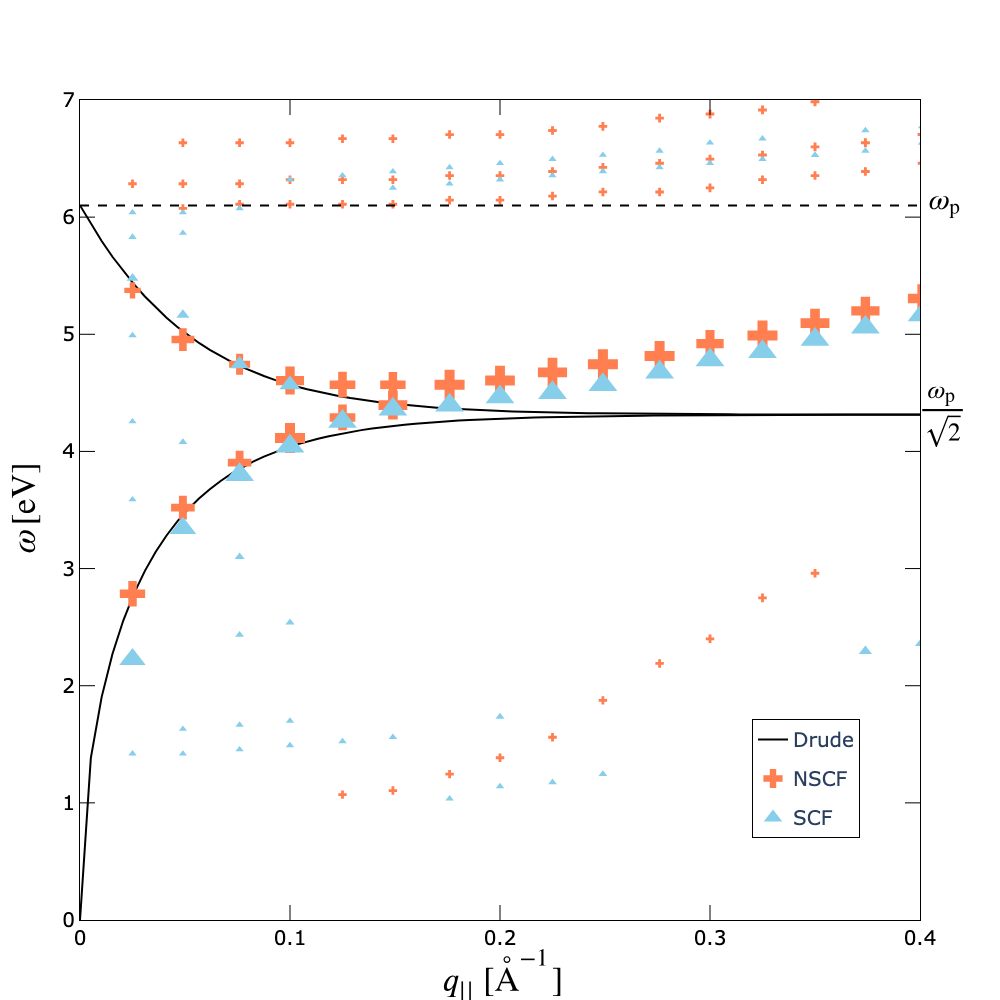} &  &
\includegraphics[width = 0.45\textwidth]{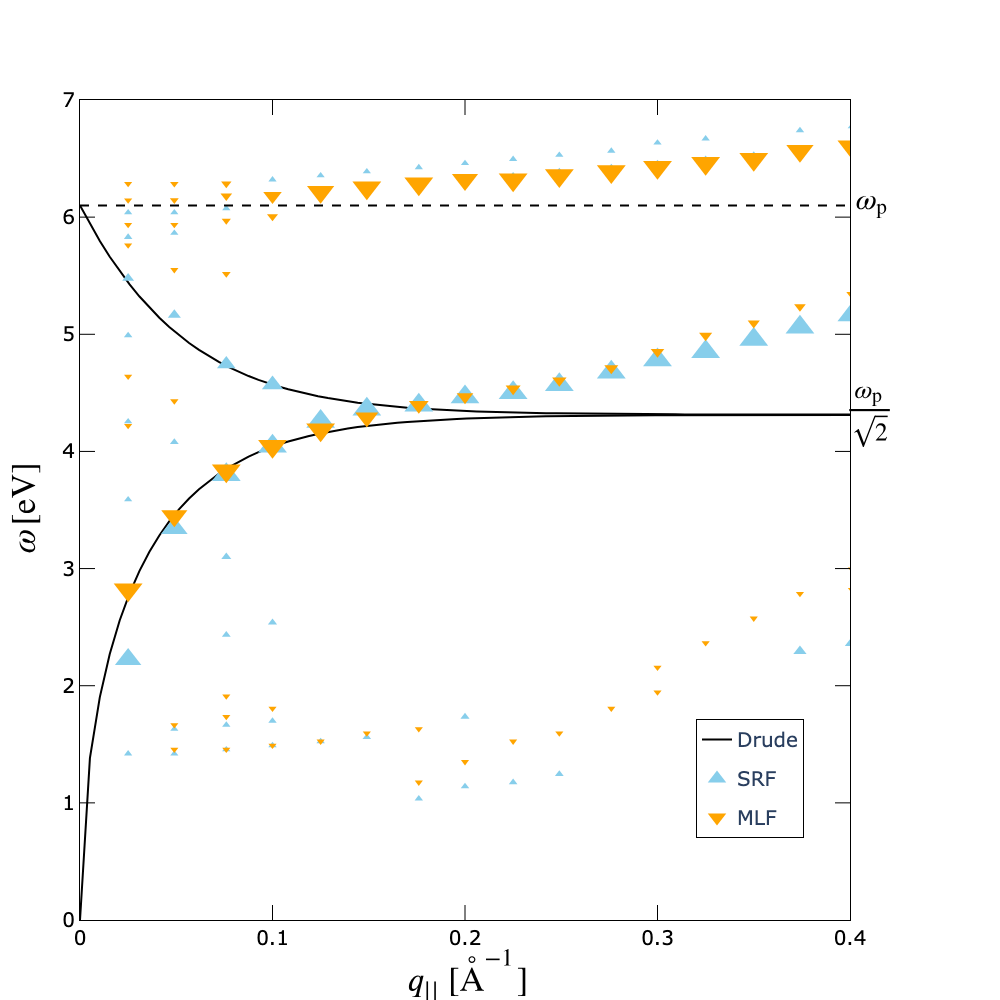}\\
\end{tabular}
\caption{Dispersion relations obtained in the jellium model (NSCF and SCF) with RPA,
compared to those obtained in the classical case with the Drude model as an approximation for $\varepsilon(\omega)$,
for a slab equivalent to a 10-layer system of Na (D = 21~\si{\angstrom}).
The size of the markers is proportional to the relative intensity of the peak with respect to the highest one of each spectrum (one per $q_{||}$). 
(a) Comparison of the peaks found in the SRFs, for the NSCF model with a potential following Eq.~(\ref{eq:NSCF}) (coral), and for the SCF model (light blue). (b)  Comparison of the peaks found in the MLF (orange) and SRF (blue), both for the SCF model.}
\label{fig:ccd}
\end{figure*}

The two density profiles (NSCF and SCF), even with different tails outside the surface, lead to the emergence of similar surface modes as already found by Echarri \textit{et al.}\cite{echarri2021optical} and in stark contrast with the hydrodynamic mode where these details matter much more\cite{ciraci2016quantum}. Figure~\ref{fig:ccd}a compares the dispersion relations obtained from the SRF for the two potentials (in the SCF case, the RPA, Eqs.~(\ref{eq:mat_eps}-\ref{eq:mat_W}), was used at the level of the response function) with those of the classical approach of dielectric continuum with a Drude model for the response function with a $\omega\ind{p} = 6.10$~eV. 
The dispersion relations show the peaks in the spectra, as illustrated in Fig.~2. 
Note that two resonances cannot be distinguished when they are very close compared to the width of the excitations. 

%\textcolor{red}{XG20240829 : is it within RPA or with the xc as well ?}
The Drude model uses a simple dielectric function\cite{maier2007plasmonics}:
\begin{equation}
    \varepsilon(\omega) = \varepsilon_0\Big(1-\frac{\omega\ind{p}^2}{\omega^2}\Big),
\end{equation}
and yields the following dispersion relation for a single slab:
\begin{equation}
    \omega_{\pm}^2(q_{||}) = \frac{1}{2}\omega\ind{p}^2\Big(1\pm e^{-q_{||}D}\Big).
    \label{Eq:38}
\end{equation}

This model gives two surface plasmons. The first one (the $-$ sign in Eq.\ref{Eq:38}) has the same symmetry as the slab, the second one (the $+$ sign in Eq.\ref{Eq:38}) 
is an antisymmetric mode.
From Fig.~\ref{fig:ccd}a, it is evident that both the SCF and the NSCF approaches result in dispersion curves very close to the dielectric model in the long wavelength limit, with a remarkable agreement up to $q_{||} = 0.15$~\si{\angstrom}$^{-1}$.
The main differences between the two approaches are  (i) the energy of the bulk mode is not well predicted by the SRF and (ii) weak electron-hole excitation continuum, lacking in the dielectric model, is observed in the jellium model. The latter appears in the Fig.~\ref{fig:ccd} and can be seen in the lower energy region ($<3$~eV) and for intermediate wavelength (0.1~\si{\angstrom}$^{-1}<q_{||}<$0.4~\si{\angstrom}$^{-1}$). 
The peaks of these excitations are however several orders of magnitude less important than the peaks of the SPP modes, leading to transfer of far less energy. We also note that the symmetric and antisymmetric SPPs merge at large momentum transfer in all the approach because of the decoupling of the surface modes due to the rapid decrease of the electric field. 

%Additionally, different tools are employed to extract plasmon response information, including SRF and MLF. 
Fig.~\ref{fig:ccd}b contrasts dispersion spectra obtained from the SRF and from the MLF.
%for a jellium slab equivalent to 10 layers of Sodium. 
%\textcolor{red}{XG20240829 : this Fig.2 should come before Fig.3 , because it shows the peaks positions and heights and thus the information that is collected in Fig.3.}
It has already been noted that the MLF approach ignores the antisymmetric mode as a consequence of the averaging associated with the macroscopic dielectric function. The relative weight of the bulk mode in the MLF dispersion is also found to increase as the wavelength is decreased. Large momentum transfers are indeed more likely to excite a bulk plasmon.

Additionally, the SCF approach leads to a lower relative intensity of the antisymmetric surface mode when compared to the NSCF one when using the SRF methodology. 

%This can also be observed in the Fig.~\ref{fig:spectra}. 
%This figure, showing the loss spectra obtained from each methodology, allows one to compare the relative intensity of each methodology, including the integral of the diagonal part of the screened interaction. 
%It can be seen that the MLF methodology does not exhibit the antisymmetric mode in the final spectra. 

%Finally, the integral of the screened interaction leads to broader peaks and highlights a lot more the bulk peak (it might be the result of several bulk modes close to one another but not resolved in the spectra). 

%\textcolor{red}{ XG20240805 : Compare this caption with the other ones, and standardize. The legend is unreadable. All font sizes should be close to the size in the normal text. Also, the thickness of the lines is too small. Lines should be plain, dashed, dotted, dashed-dotted lines in addition to being colored.}
\begin{figure*}
\begin{tabular}{cccc}
(a) & & (b) & \\
& \includegraphics[width = 0.45\textwidth]{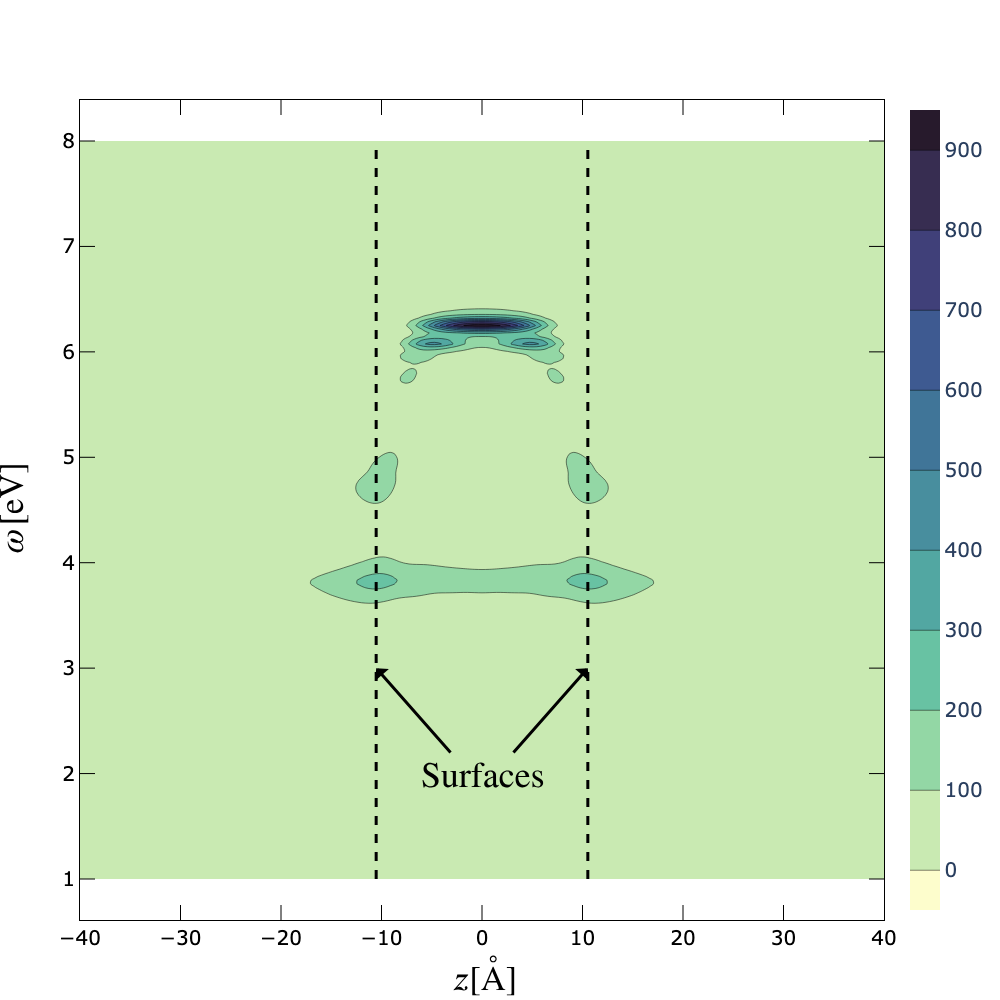} &  &
\includegraphics[width = 0.45\textwidth]{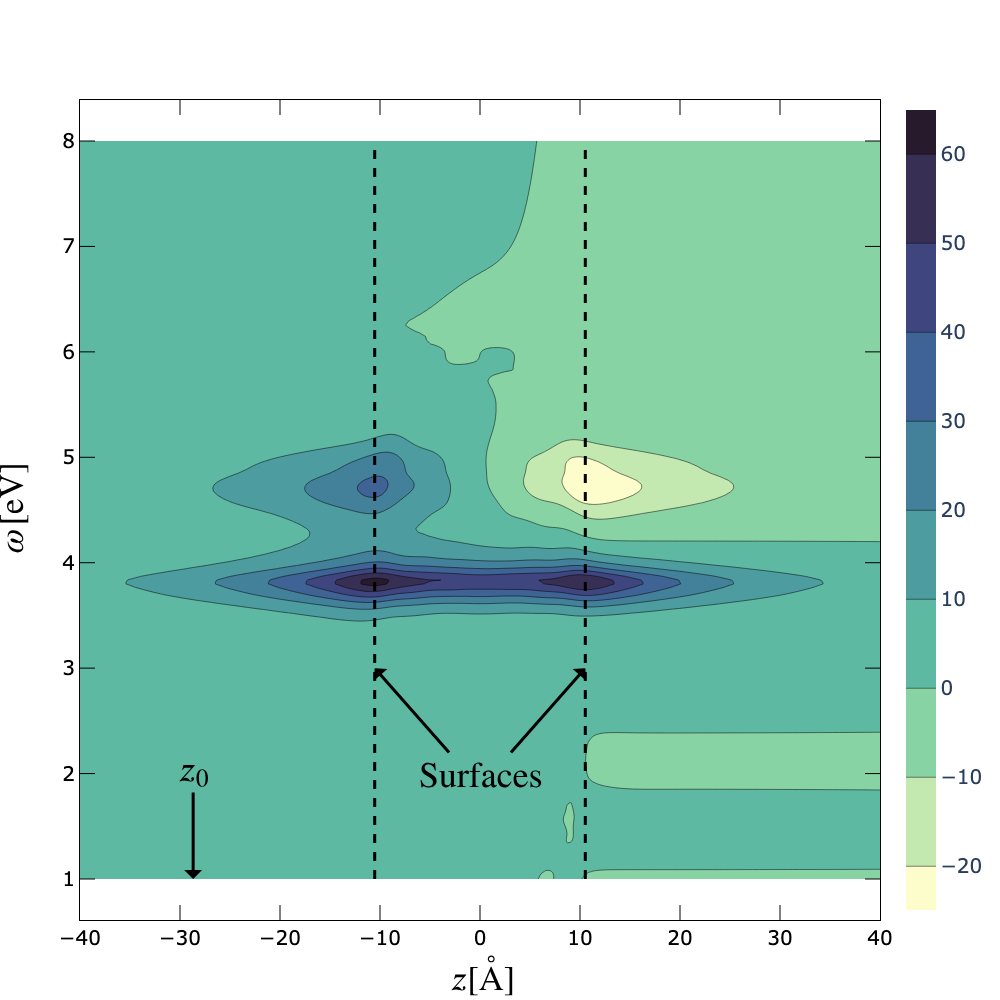}\\
\end{tabular}
\caption{(a) $\text{Im}\,W(z,z;q_{||}, \omega)$ for $q_{||} = 0.076$~\si{\angstrom}$^{-1}$. Two surface modes are visible: the symmetric one is at $\omega = 3.83$~eV and the antisymmetric one at 4.71~eV. A bulk plasmon appears at 6.16~eV. (b) $\text{Im}\,W(z_0 ,z;q_{||}, \omega)$ for $z_0 = -13.62$~\si{\angstrom} and $q_{||} = 0.076$~\si{\angstrom}$^{-1}$. In this case, the symmetry of the SPP is clearly visible.}
\label{fig:scr_int_ill}
\end{figure*}
The comparison of Fig.~\ref{fig:ccd} to the results gathered by Andersen~\textit{et al.} 
(see Fig.~5 of Ref.~\onlinecite{andersen2012spatially}), 
for the same system but with first-principles methods, highlights the ability of the jellium model to identify the symmetric surface plasmon.
However, the antisymmetric surface plasmon predicted in the jellium case is found to follow less closely the first-principles results and more closely the dielectric model in the long wavelength (small $q_{||}$) regime (the first-principles results present a strong redshift, up to 1~eV, for the antisymmetric mode with respect to the classical and our jellium models). It is however expected that in this limit, classical and quantum methods yield the same results. This results probably from the smaller amount of vacuum between periodic replicas of the metallic film in the supercell approach in the first-principle calculations. In the jellium case, we found the symmetric mode harder to converge with respect to the amount of vacuum than the antisymmetric one. Yet, thanks to the smaller computational requirements, a better convergence with respect to this parameter can be performed for the two SPP modes in the jellium case.  

In the short wave-length regime (large $q_{||}$)  on the other hand, non-local effects become significant and the quantum results prevail. Our jellium calculations show the same $q_{||}^2$ dispersion as the one found in the first-principle data~\cite{andersen2012spatially}\cite{6}. 
This highlights capabilities of the jellium model to correctly describe the quantum effects in both the long and short wave-length regimes, as long as non-local effect associated with the atomistic description of the problem is not too important (which is not the case for Na).

Spatial information can be gleaned from the imaginary part of the screened interaction, as demonstrated in Fig.~\ref{fig:scr_int_ill} for  $q_{||} = 0.076$~\si{\angstrom}$^{-1}$.
the left part of the Figfre, the function is analyzed forAna response at the perturbation location ($z = z')$, as suggested from the EELS theory. Although two surface modes are clearly observed at $\omega = 3.83$~eV and at $\omega = 4.71$~eV , along with a bulk plasmon at $\omega = 6.18$~eV, the symmetries of the SPP modes cannot be resolved from the figure. 
This could be alleviated by setting the perturbation location on one side of the surface and analyzing the response over the whole range of position ($z\neq z'$) as shown in the right part of the Fig.~\ref{fig:scr_int_ill} for $z_0 = -13.62$~\si{\angstrom}. In this case, the bulk mode is not excited but the symmetry of the surface mode is revealed. 
The symmetric mode appears at 3.83~eV while the response is antisymmetric for the 4.71~eV mode. 
Since it is more complex to extract all the relevant information from the screened interaction and it does not bring any added value, the following discussion will focus on the MLF and SRF.

Overall, these results underscore the interest of the jellium model and the possibility to obtain in depth results by combining different methodologies. 
Even the absence of a mode in the MLF helps to characterize its symmetry. 
Its importance will prevail even more when moving to more complex systems, when the visualization of the symmetries thanks to the imaginary part of the screened interaction will be more cumbersome to get.

\subsection{Double slab}

In this section, the analysis is extended to double-slab systems spaced by \(d\ind{gap}\): they are composed of two jellium slabs equivalent to Na slabs of 6 atomic layer (12.6~\si{\angstrom}), see Fig.~\ref{fig:potential}c. 
The dispersion curves are examined with respect to both \(d\ind{gap}\)
and with respect to the parallel wavevector.

Before exploring the results for the jellium model, we introduce the classical results obtained by Richter \textit{et al.} for similar slab systems\cite{richter1981energy}. 
For a layer of dielectric function $\varepsilon_2 $ and thickness $d\ind{gap}$ sandwiched between two layers of dielectric function $\varepsilon_1$ and thickness $d_1$, they obtained the following dispersion relation for the symmetric and the antisymmetric surface modes\cite{7}:
\begin{widetext}
\begin{equation}\label{richter1}
 \varepsilon_1\Big[(1+\varepsilon_1)\exp(q_{||}d_1)+(1-\varepsilon_1)\exp(-q_{||}d_1)\Big]\tanh\Big(\frac{q_{||}d\ind{gap}}{2}\Big)+
    \varepsilon_2\Big[(1+\varepsilon_1)\exp(q_{||}d_1)-(1-\varepsilon_1)\exp(-q_{||}d_1)\Big] = 0,
\end{equation}
and:
\begin{equation}\label{richter2}
\varepsilon_1\Big[(1+\varepsilon_1)\exp(q_{||}d_1)+(1-\varepsilon_1)\exp(-q_{||}d_1)\Big]\coth\Big(\frac{q_{||}d\ind{gap}}{2}\Big)+
    \varepsilon_2\Big[(1+\varepsilon_1)\exp(q_{||}d_1)-(1-\varepsilon_1)\exp(-q_{||}d_1)\Big] = 0.
\end{equation} 
\end{widetext}
In our case, $\varepsilon_2 = 1$, since vacuum separates the two slabs.

\begin{figure}
\includegraphics[width = 0.5\textwidth]{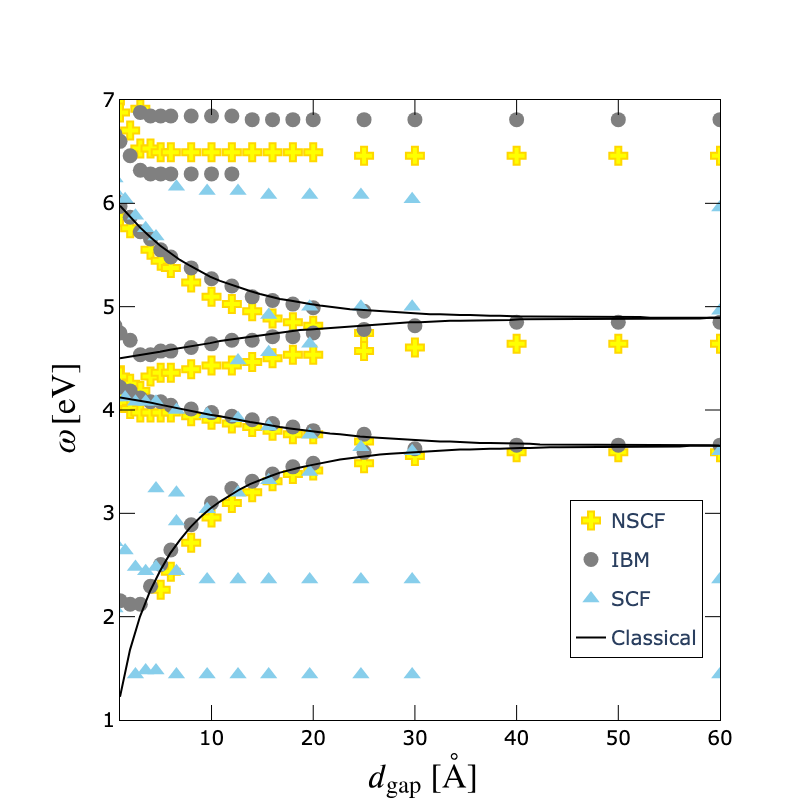}
\caption{Evolution of the plasmon frequency found in the SCF case for the SRF in the double-slab system (each slab with a width of 12.6~\si{\angstrom}) as the gap size is tuned. $q_{||} = 0.076$~\si{\angstrom}$^{-1}$.
The crosses represent the NSCF case, the triangles, the SCF case and the dots, the IBM approximation (NSCF with C = $\infty$). 
The black lines correspond to the classical approximation where the Drude model is used for the dielectric function. 
}
\label{fig:gap_size}
\end{figure}

Figure~\ref{fig:gap_size} presents the evolution of the plasmonic modes as a function of the distance between the layer, $d\ind{gap}$, for $q_{||} = 0.076$~\si{\angstrom}$^{-1}$. Classical  (in the Drude approximation, $\omega\ind{p}= 6.1~eV$) and jellium results present similar trends, with four distinctive surface modes merging two by two as $d\ind{gap}$ increases. 
For large \(d\ind{gap}\), i.e. when the slabs are well separated, the modes obviously akin to those in isolated slabs are recovered. 
Conversely, as \(d\ind{gap}\) decreases to 0, the excitations are those of a single slab with a doubled width with two surface modes at $\approx 4.1$~eV and $\approx 4.5$~eV and the bulk mode at 6.1~eV.  
Within an intermediate range of \(d\ind{gap}\) ($4$~\si{\angstrom}  to $45$~\si{\angstrom}), new surface modes emerge. 
They are particularly noticeable at small parallel wavevectors ($q_{||} <$ 0.25~\si{\angstrom}$^{-1}$) as will be analyzed afterwards. For the time being, we continue the analysis for $q_{||} = 0.076$~\si{\angstrom}$^{-1}$.

We note that the IBM model is the closest to the classical approximation. 
The NSCF model is globally red-shifted by about 0.15~eV with respect to the latter (it can be seen on Fig.~\ref{fig:potential} that the density profiles for the single and double slab scenarios are different, implying that the tuning of the barrier height is no longer valid to reproduce the SCF results). 
Tuning the height of the barrier in the NSCF potential shows that, as its height decreases, the SPP frequencies are continuously red-shifted. 
This can be linked to the $\omega\ind{p}$ value: as the size of the barrier is decreased, the electron gas is more spread, leading to a smaller average electronic density and hence, to a smaller plasma frequency. 
The SCF model yields generally peaks close to the classical approximation. Nonetheless, some discrepancies can be found in the lower energy mode and in the two higher energy modes (the bulk modes are not analyzed here). 
In the former, the differences for very small gaps ($d\ind{gap}<7$~\si{\angstrom}) might be attributed to tunneling effects (in the NSCF case, the definition of the gap is more ambiguous as the parameter $d$ becomes larger than $d\ind{gap}$ and so no conclusion could be made for $d\ind{gap}$ smaller than 3~\si{\angstrom}). 
In the two higher energy modes, some peaks, below $d\ind{gap}=12.61$~\si{\angstrom} for the lowest one and between $d\ind{gap}=$4.54 and 15.62~\si{\angstrom} for the highest one, are not resolved as they are hidden in the main SPP peak of the second lowest energy mode. 
Some peaks between $d\ind{gap} = 12.61$ and 24.69~\si{\angstrom} are also red-shifted in the two higher energy modes with respect to the classical approximation. 
Those are hence better reproduced by the NSCF model than by the IBM one.
It is important to note that the red-shift of the NSCF model does not impact the description of the modes and the trends are still very similar in the SCF case and NSCF one.  
In the lowest part of the graph, one can observe two horizontal lines which are associated to weak electron-hole excitations. The rest of the discussion in this section is based on the SCF results.  

%The surface modes are split in this small \(q_{||}\) regime, a phenomenon absent in large $d\ind{gap}$ and very small $d\ind{gap}$ regimes. %It is easily understood as when the gap becomes very small, the jellium behaves as one slab while when the gap becomes very large, the two slabs do not interact with one another.
 
\begin{figure*}
\includegraphics[scale=0.5]{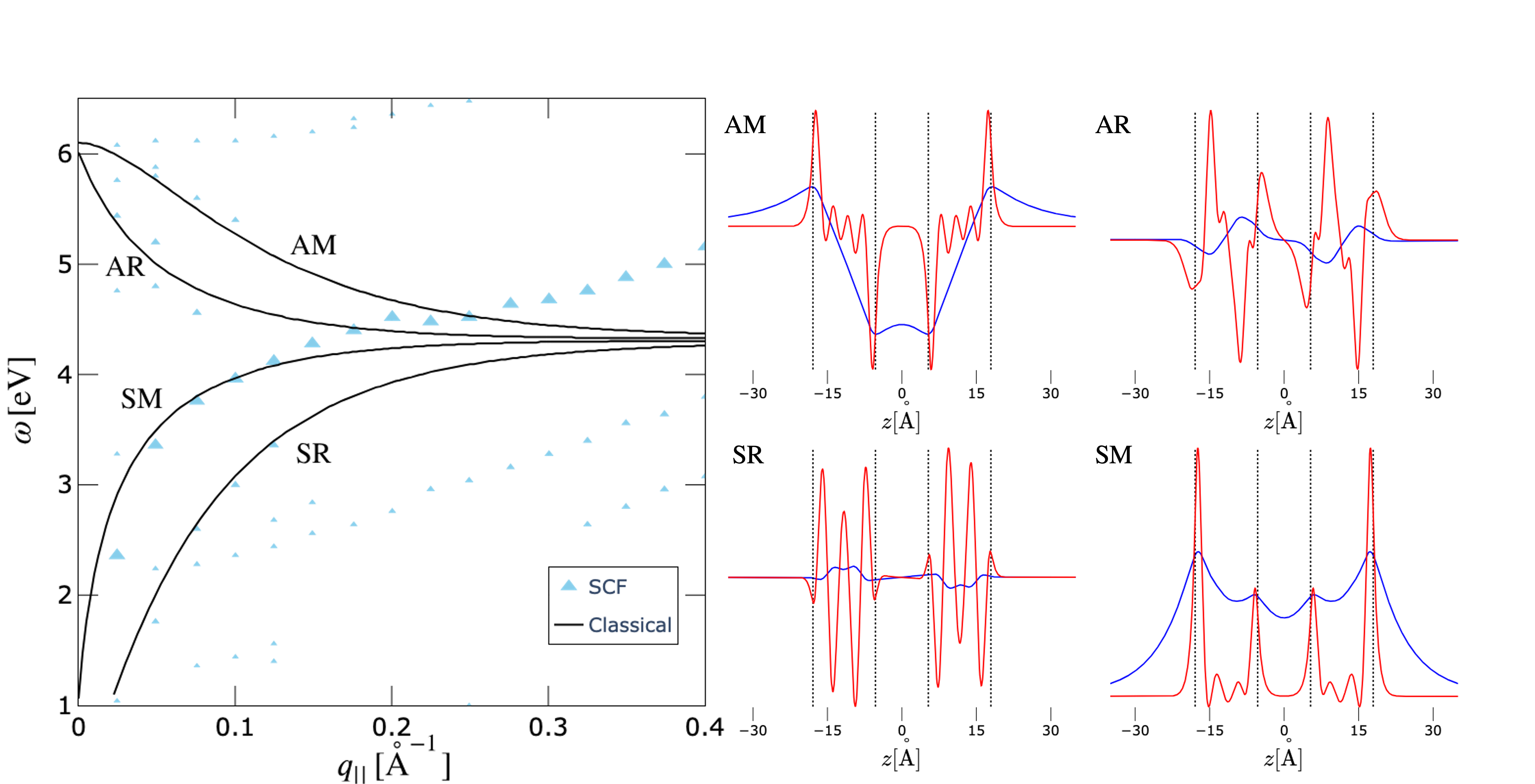}
\caption{Dispersion relation of the double symmetric 6-atomic layer Na equivalent slab system with a gap of 9.82~\si{\angstrom} using the SRF. 
Left: In black, the prediction of the classical theory from Eqs.(~\ref{richter1}) and~(\ref{richter2}). 
In blue, the prediction from the SCF. The size of the dots is proportional to the intensity of the mode (relative intensity for each $q_{||}$). 
Right: the 4 modes associated with each branch with the typical shape of the variation of, in red, the density, $\rho_i(z)$ and, in blue, of the potential, $V_i(z)$ associated with the mode. 
Top-left is the AM mode, top-right is the AR mode, bottom-left is the SR mode and bottom-right is the SM mode. }
\label{fig:mode_2slabs}
\end{figure*}

The behavior for different parallel wavevectors is illustrated in Fig.~\ref{fig:mode_2slabs}. 
The figure displays a typical dispersion curve for two 6-layer Na equivalent slabs with a gap of 9.82~\si{\angstrom} for the classical dielectric approach and for the jellium model in the SRF-SCF approach.  
Four surface modes and one bulk mode are identified together with the electron-hole excitations. 
The bulk plasmons and the electron-hole excitations have very little weight and exhibit minimal deviations from the single-slab case, as expected. 

For further characterization of the surface-plasmon modes, two types of symmetries are used in this analysis. As described previously, for one single slab, a plasmon mode can be symmetric (S) or antisymmetric (A).
Considering now one slab in interaction with another slab, such symmetry is broken, although it might hold approximately. 
Such a local symmetry will be called ``inner symmetry".
By contrast, the ``global symmetry" is the true symmetry of the double-slab system, one slab being the mirror of the other.
The modes will be labelled as mirror (M) ones when they are globally symmetric and reversed (R) ones when they are globally antisymmetric. 
The symmetry of each mode will be identified thanks to the density change obtained by the decomposition in the MLF (see right panel of Fig.~\ref{fig:mode_2slabs}).
As expected from the analysis of the single-slab case, reversed modes are not found in the MLF as they are globally antisymmetric. 

We examine now the left panel of
Fig.~\ref{fig:mode_2slabs}).
Starting from the lowest frequency, the branch labeled SR
corresponds to a mode being inner symmetric and having a global antisymmetry from one slab to the other (SR mode). 
The second branch, SM , corresponds to a (roughly) locally symmetric mode mirrored from one slab to the other (doubly symmetric, SM mode). 
The third branch, AR, is an antisymmetric mode reversed (AR mode) and the highest branch, AM, is an antisymmetric mode mirrored (AM mode). 

Comparing the dispersion curves computed here with the ones of the classical model, the general trends are similar but the agreement is less good than for a single slab in the small wavevector regime. 
The SM and SR modes follow the classical approximation while the AM mode is slightly blue shifted and the AR mode is not well resolved.
In the large wavevector regime, the same discrepancy is found as in the single-slab model where the four modes become degenerate because of the decoupling between the four surfaces in both approach. 
In the jellium model, a quadratic dispersion relation is observed whereas the dispersion curve flattens in the classical case to tend towards $\omega\ind{p}/\sqrt{2}$.

%, only the SM and AM modes are found in the spectra.

Analyzing the predicted intensity of these modes reveals that the SM mode is the most intense, followed by the AR mode.
The two remaining modes exhibit similar intensities, significantly lower than the AR mode. Comparing these observations with the classical case, it can be concluded that the SR mode and AM mode intensities are too small in very small gap or very large gap regime to be observed, despite being still expected from the classical theory. 
Note however that this analysis might change if another response function characterizing a different experimental setup is used.

\begin{figure}
\includegraphics[width = 0.5\textwidth]{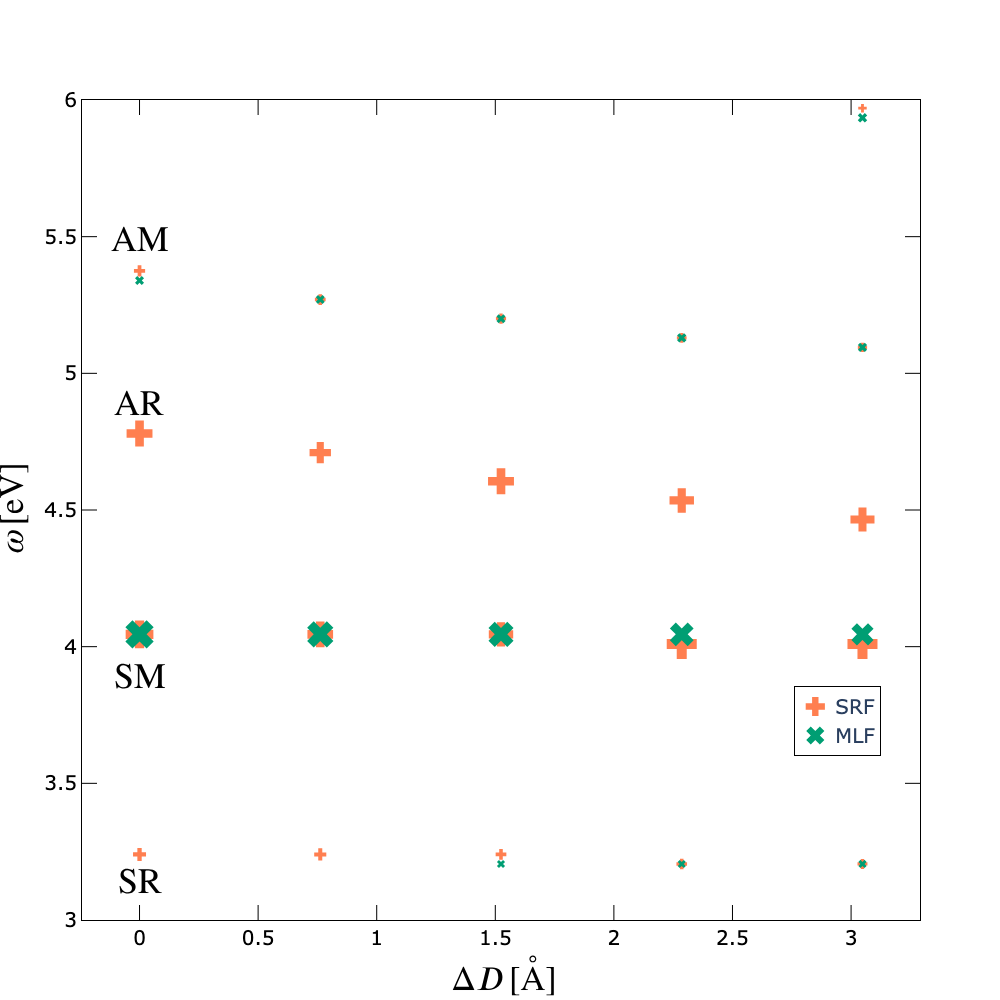}
\caption{Evolution of the surface mode frequency with respect to $\Delta D$, the difference of width between the two slabs. The width of the left slab is fixed at 12.6~\si{\angstrom}. The coral straight crosses present the results for the SRF while tilted green crosses show the results for the MLF. The size of the symbols is proportional to the relative intensity of the peaks for a given $\Delta V$.}

\label{fig:deltaL}
\end{figure}

Finally, in anticipation to the case of two different slabs treated in the next section, it can be noted that the four modes still appear in case where the two slabs do not have the same width (Fig.~\ref{fig:deltaL}). 
The SM and SR modes do not present any significant changes in the resonant frequency when increasing the size of one of the two slabs by up to 3.18~\si{\angstrom}, keeping the same $d\ind{gap} = 9.82$~\si{\angstrom}. 
The AM and AR modes are still present but are red shifted by about 0.2~eV when the size is increased by 3.18~\si{\angstrom}, see Fig.~\ref{fig:deltaL}. 
%On the other hand, if the two slabs are made from two distinct materials (here represented by two distinct jellium densities), the picture changes progressively. The analysis of this case is made in the next section.

\subsection{Two different slabs}

The case of two slabs made of different materials (i.e. with different densities) is now investigated. 
The following analysis considers only the NSCF model. 
As concerns the dielectric response, in the two previous sections this NSCF model was seen not to differ fundamentally from the SCF one, reproducing the trends and the principal modes accurately.
Moreover, analyses of the trends are easier with the NSCF model as the computation time is decreased since the SCF part is bypassed,
while, in our numerical investigations, we found the stabilization of the SCF ground state to be problematic in some cases.

The investigated systems are made of two slabs with a distance $d\ind{gap} = 10.64$~\si{\angstrom} with one of the well deeper than the other. The total number of electrons in the system being kept constant, and one of the well being a preferential place for the electrons, an asymmetric repartition naturally arises. It results in slabs of two different materials that can then be defined by their electronic densities (not known beforehand) and, since the neutrality is imposed, their background density can be computed \textit{a posteriori}.
The analysis will focus on the following aspects: the impact of the ratio of density between the two slabs and the dispersion relations obtained from the MLF and SRF.

\begin{figure}
\includegraphics[width=0.45\textwidth]{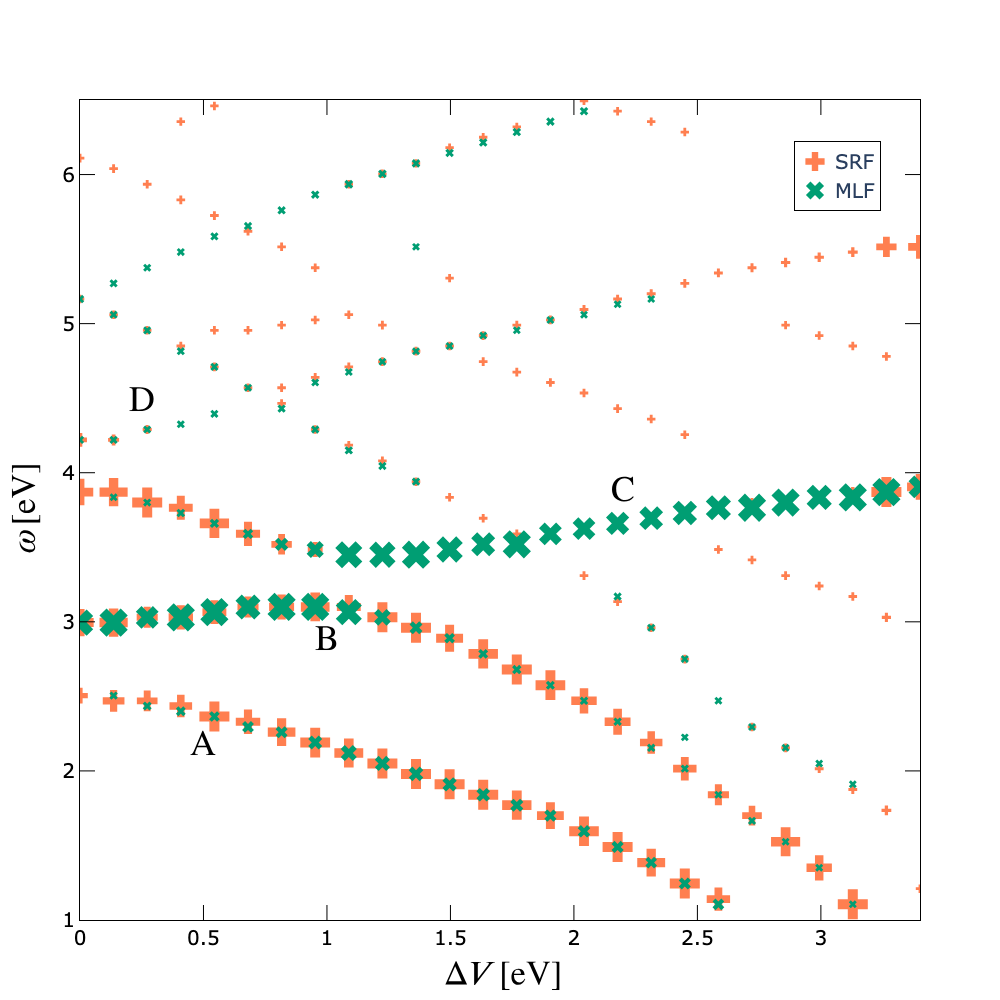}
\caption{Evolution
of the plasmon modes for the two-different-slab system with respect to the difference of potential between the two wells in the NSCF potential profile. The left well is kept at a fixed energy ($-13.61$~eV, the vacuum being at 0), while the right well is lifted by $\Delta V$.  The coral straight crosses present the results for the SRF while tilted green crosses show the results for the MLF. The size of the symbols is proportional to the relative intensity of the peaks for a given $\Delta V$.}
\label{fig:asym_l}
\end{figure}

\begin{figure}
\includegraphics[width=0.45\textwidth]{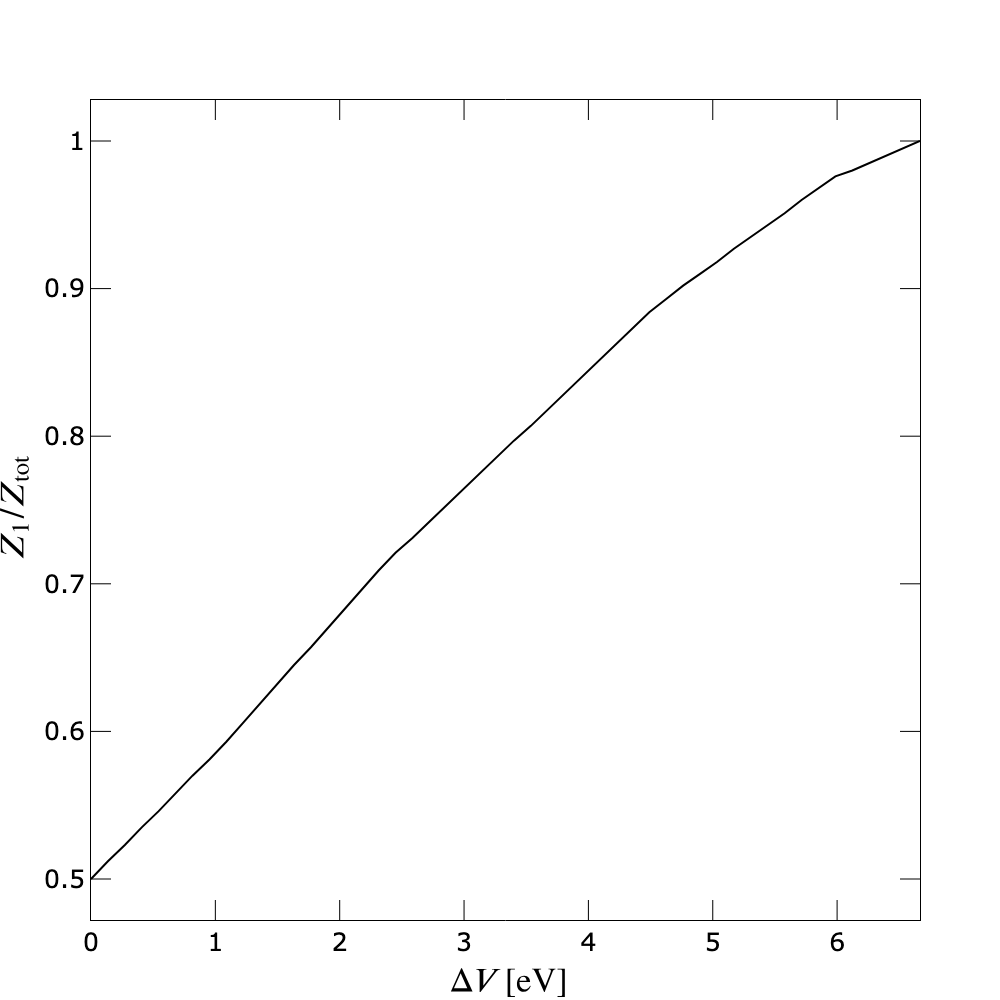}   
    
\caption{Ratio of the amount of electrons in the left slab  and the total amount of electrons in the system when $\Delta V$ is tuned.} 
\label{fig:ratiovsdeltaV}
\end{figure}
\begin{figure*}  
    \includegraphics[width=0.9\linewidth]{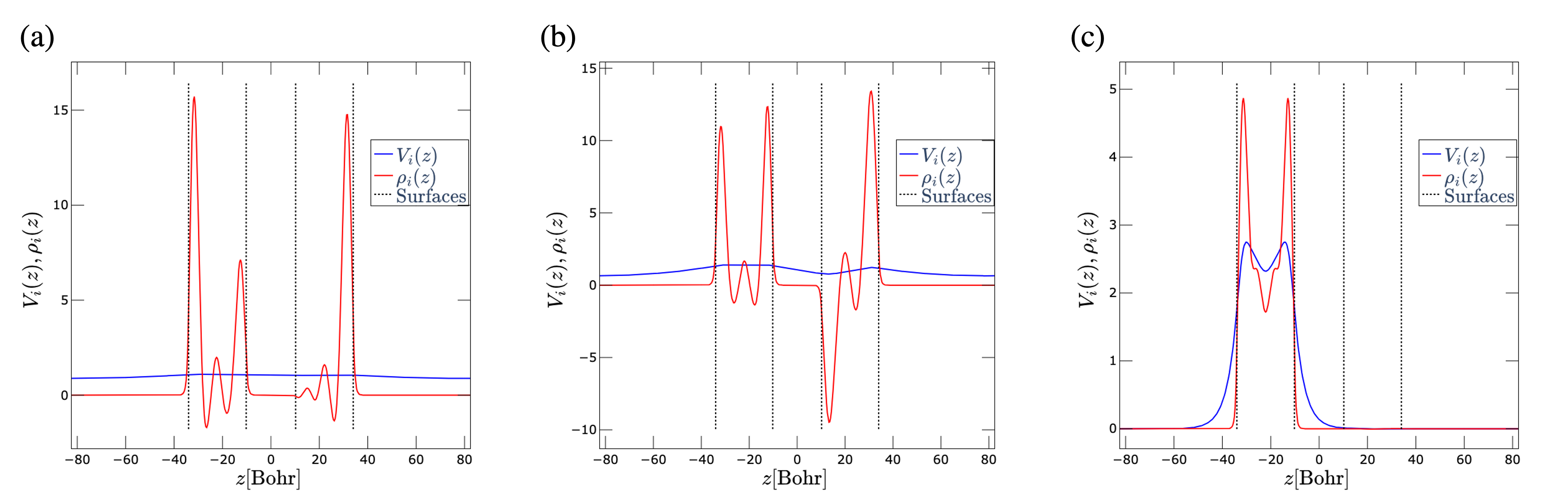}   
    
\caption{Surface modes of the branch (F) when $\Delta V = 0.41$~eV. (a)  Example of a SR mode for $q_{||} < 0.1$~\si{\angstrom}$^{-1}$. (b) Example of a mode exhibiting symmetric-antisymmetric coupling at $0.1< q_{||} < 0.3$~\si{\angstrom}$^{-1}$. (c) Only the mode in the left slab is excited in the branch (F) when $q_{||} > 0.3$~\si{\angstrom}$^{-1}$} 
\label{fig:modes}
\end{figure*}

Fig.~\ref{fig:asym_l} presents the evolution of the surface plasmon gathered with the SRF when the electron density of the two slabs is tuned. 
The left-well depth is kept at 0.5~Ha while the right-well one is raised by $\Delta V$ with respect to the left well. The $\Delta V=0$, leftmost value in the figure, corresponds therefore to identical slabs coupled. The parallel wavevector is set to 0.1~\si{\angstrom}$^{-1}$.

The rightmost position shown corresponds to a case where the difference in energy is such that the left slab contains almost all the electrons, the right one being almost completely depleted. 
The Fig.~\ref{fig:ratiovsdeltaV} shows the correspondence between the ratio of electrons in the left slab and the total number of electrons in the system as a function of the difference of the potential.

The graph presented in Fig.~\ref{fig:asym_l} can be divided in three zones. In the first one, until $\Delta V \approx 0.5$~eV the SPP modes can be seen as perturbations of the perfectly symmetric case. 
It is exemplified in Fig.~\ref{fig:modes}a where the modes share similar characteristics with the SM mode of Fig.~\ref{fig:mode_2slabs} with only the inner density peaks of the right slab being less pronounced. 
The second zone extends up to $\Delta V \approx 2$~eV. 
The difference in density compared to the symmetric case is more pronounced, presenting a coupling of one symmetric mode with one asymmetric mode to form the branches labeled (B) and (C).  
The resulting density is shown in Fig.~\ref{fig:modes}b. 
For the other two branches, labeled (A) and (D), the coupling between symmetric or antisymmetric modes is weaker. 
Finally, for $\Delta V>2$~eV, the modes of the two slabs become strongly decoupled because the left slab has taken up more than 70\% of the electron density. 
The frequencies of the modes of the right slab (A and B branches, Fig.~\ref{fig:modes}c) then rapidly decrease as the electron density drops.
Interestingly enough, the presence of a second slab makes the branch that starts with the SM mode, the branch (B), go up to reach a maximum at around $\Delta V = 0.82$~eV.

When looking in detail at the interplay between the modes, taking into account their spatial description, the situation where the two symmetric modes are coupled to form a SM mode, branch (B), is quickly (at about $\Delta V = 0.5$~eV) replaced by a situation where the antisymmetric mode of the right slab couples with the symmetric mode of the left slab (which contains more and more electrons as $\Delta V$ increases). 
After the maximum in (B), the spatial description shows a fading of the coupling, until the antisymmetric mode of the right slab is the only one excited. 
The branch (A), the one starting with the SR mode, presents a negative dispersion, and the symmetric mode of the right slab only couples lightly with the symmetric mode of the left slab. 

The third branch, labeled (C), the one that starts with the AR mode and finishes at the position of the symmetric mode of the left slab, presents somehow the same features as the branch (B). 
The antisymmetric mode of the right slab is the main feature until the antisymmetric mode of the left slab finishes to be transformed in a symmetric mode.
Then, the latter becomes more and more the dominant mode until the antisymmetric mode of the right slab does not couple with it anymore. 
Finally, the upper branch, labeled (D), in the same way as the branch (A), starts with the AM mode but the antisymmetric mode of the left slab quickly becomes the leading mode while the coupling with the antisymmetric mode of the right slab becomes less prevalent before fading completely when the density becomes almost null in this slab.

The dispersion relations for different fixed $\Delta V$ are now analyzed. Just as in the dispersion with respect to variation of the potential, we observe that the SPP modes evolve along each branch. The possibility to couple or not with another plasmon mode makes the mode at the small wavevector very different from the one observed at large wavevector of the same branch. Again, one can distinguish three zones in the dispersion relations. A first zone at long wavelength where the symmetries of the identical slabs case prevail. This zone becomes very narrow when $\Delta V$ increases. A second zone where the coupling between symmetric and antisymmetric modes are possible. This zone also shortens with the increase of $\Delta V$. Finally, at very short wavelength, the modes are decoupled even for small differences of electronic density. We call these zones Zone I, Zone II and Zone III, respectively.
From the comments made above, it is clear that for $\Delta V$ larger than $2$~eV, the SPP of two slabs are decoupled and the dispersion curves are the superposition of the curves of the two isolated slabs. Indeed, they occurs at very different energy due to the density difference. 

\begin{figure*}
\begin{tabular}{cccc}
(a) & & (b) & \\
& \includegraphics[width=0.45\linewidth]{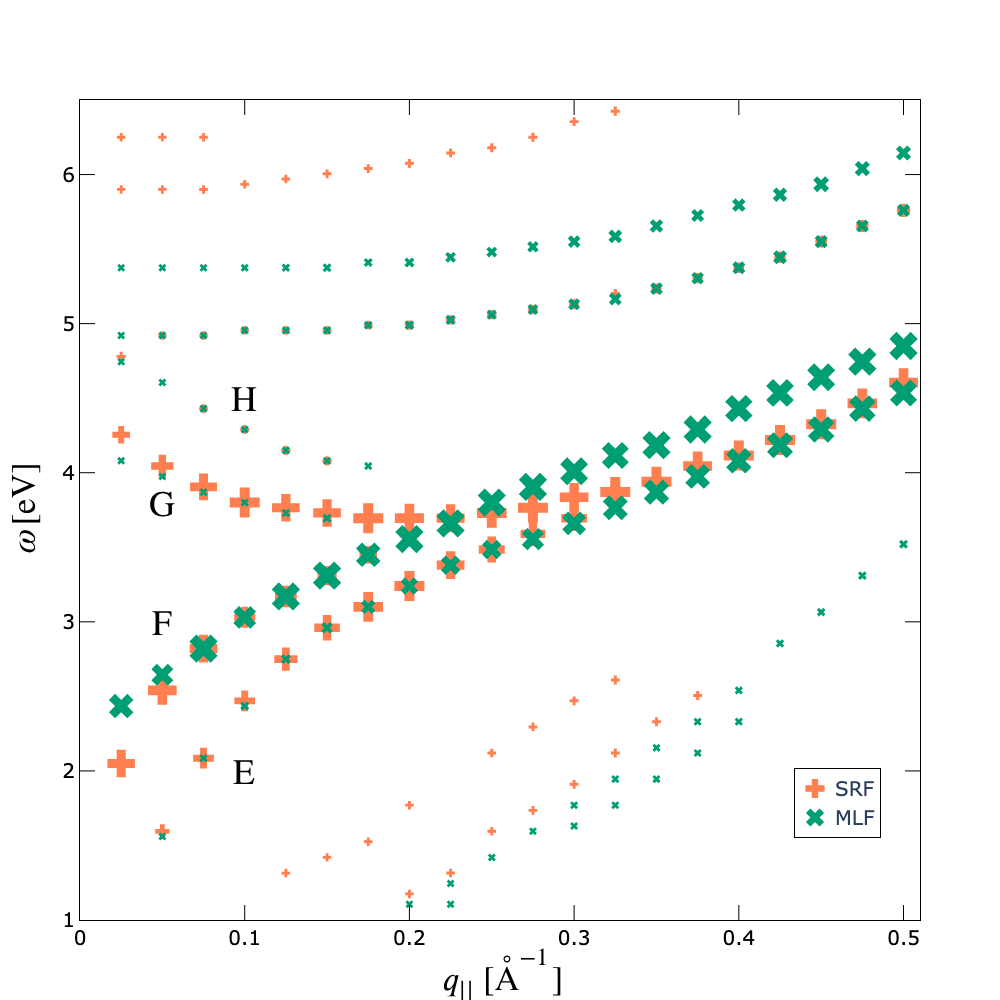} &  &
\includegraphics[width=0.45\linewidth]{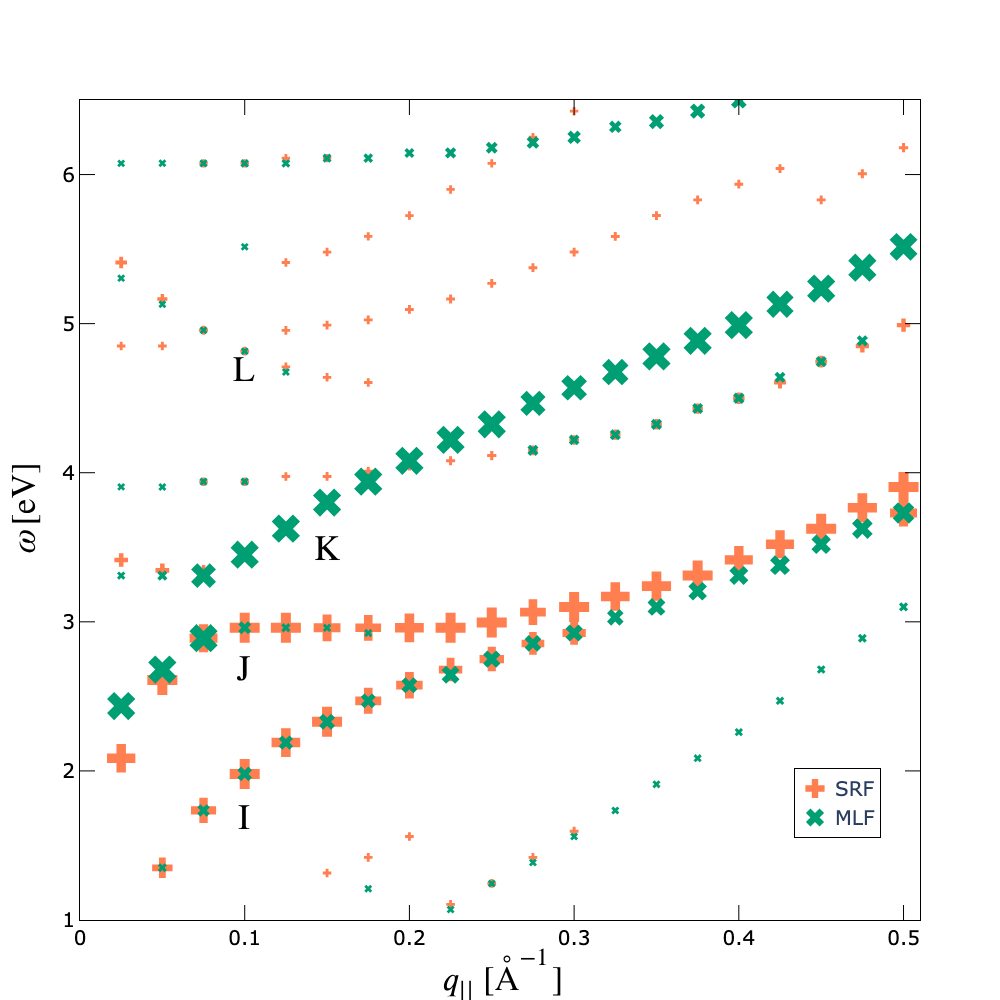}\\
\end{tabular}
\caption{Dispersion relation of the two-different-slabs system using the NSCF approach with the results for the MLF in green and for the SRF in coral. (a) $\Delta V = 0.41$~eV case (53.5\% of the electrons in the left slab). (b) $\Delta V = 1.36$~eV case (62\% of the electrons in the left slab).}
\label{fig:asym_qp_l}
\end{figure*}

Before this value, the densities are relatively close, allowing the SPPs of the two slabs taken separately to occurs at similar energy, leading to a possible coupling.  
Figure~\ref{fig:asym_qp_l} shows the dispersion curves for $\Delta V = 0.41$~eV (53.5\%-46.5\% electron distribution) and $\Delta V = 1.36$~eV (62\%-38\%).

In the $\Delta V = 0.41$~eV scenario, Zone I goes up to $q_{||} \approx 0.06$~\si{\angstrom}$^{-1}$, Zone II finishes around $q_{||} \approx 0.3$~\si{\angstrom}$^{-1}$ and Zone III is everything above this value. Four main branches labeled (E), (F), (G) and (H) are distinguished in Fig.~\ref{fig:asym_qp_l}a. In the MLF methodology, branches (G) and (H) merges with branch (F) in the around $q_{||} \approx 0.2$~\si{\angstrom}$^{-1}$. In the SRF approach, the branch (H) first merge with the (G) and (F) at $q_{||} \approx 0.2$~\si{\angstrom}$^{-1}$. They then merge altogether with branch (E) at $q_{||} \approx 0.3$~\si{\angstrom}$^{-1}$.

As mentioned above, in Zone I, the four modes are slight variations of the ones found in the symmetric case (see Fig.~\ref{fig:modes}a) and the SRF and MLF approach give the same results. As $q_{||}$ increases, some mode starts to fade. In the branch (E), its is the symmetric mode of the left slab. In branch (F), it is symmetric mode of the right slab. In branch (G) and (H) it is the antisymmetric mode of the left and right slab respectively.

In Zone II, branch (E) and (G) continue the transformation started in Zone I. The fading modes are now almost negligible and the SPP mode is driven by the symmetric mode of the right slab in branch (E) and by the antisymmetric mode of the left slab in branch (H). In the branch (G), the transformation continues. The antisymmetric mode of the left slab remains the same while the antisymmetric mode of the left slab is transformed. The transformation affects primarily the rightmost part of the mode, changing gradually the sign of the variation from an increase of density to a decrease in the inner part of the slab. This transforms the mode from an antisymmetric mode to a symmetric one. The branch (G) corresponds to a symmetric antisymmetric coupling. The branch (F) presents a similar transformation. The symmetric mode of the right slab becomes antisymmetric as the variation of density on the innermost part of the slab gradually changes sign. The merge occurs when the two branches (F) and (G) have fully transited from their original coupling to a symmetric-antisymmetric coupling. In the MLF picture, this coupling remains after branches the merge (even with branch (H) merging) but as $q_{||}$ increases, the antisymmetric mode starts fading. In the SRF picture, it is the symmetric mode that fades, leaving the antisymmetric mode the most prominent one before this branch merges with the branch (E). This merge corresponds is very similar with what happens in the single slab scenario where the energies of antisymmetric and symmetric mode (of the right slab here) become degenerate. Hence, the symmetric mode prevails after the merge.

In Zone III, the branch (E) corresponds to the dispersion of symmetric SPP of the right slab and the branch (F) to the symmetric SPP of the left slab (shown in Fig.~\ref{fig:modes}c). The branch (F) is absent in the SRF methodology. The reason is that the SRF modelize a $z>0$ electron trajectory so that the mode of the slab located in $z<0$ has a vanishing weight. Note that, just as in the single slab case, the antisymmetric mode is absent at short wavelength in both methodologies.

In the case where $\Delta V = 1.36$~eV, presented in Fig.~\ref{fig:asym_qp_l}b, Zone I and II are shorter, with their limits at $q_{||} \approx 0.02$~\si{\angstrom}$^{-1}$ and $q_{||} \approx 0.13$~\si{\angstrom}$^{-1}$, respectively. The analysis of the mode is relatively similar to the one of the case $\Delta V = 0.41$~eV. Again four branches are highlighted, labeled (I), (J), (K) and (L) in Fig.~\ref{fig:asym_qp_l}b.

The branch (I) is equivalent to the branch (E), the transition from the initial state to the final one as $q{||}$ increases being faster in this case. 
The branch (L) is also similar to branch (H). The starting modes of branch (J) and (K) (SM and AR) quickly transition to a state of symmetric-antisymmetric coupling but rather than merging in this case, they avoid crossing ($q_{||} = 0.04$~\si{\angstrom}$^{-1}$ with the antisymmetric mode becoming the leading mode of branch (J) while the symmetric mode is more important in branch (K). 
Two distinct merging occurs then, one at $q_{||} = 0.13$~\si{\angstrom}$^{-1}$ between the branch (K) and the branch (L) and one at $q_{||} = 0.17$~\si{\angstrom}$^{-1}$. Those merges are again equivalent to the ones of the single slab case as the two slabs are decoupled. 
In Fig.~\ref{fig:asym_qp_l}b, a branch starting at around 3.90~eV can be observed. It corresponds to the bulk mode of the right slab. As the bulk mode does not produce an induced field large enough outside the slab, it can not couple with the other slab. 
Finally, one can observe two differences between the SRF and MLF methodologies. The first one is the absence of the branch (K) in the SRF approach, for the same reason as in the $\Delta V = 0.41$~eV. Then, the energy prediction for the mode of branch (I) is slightly blue-shifted in the SRF approach with respect to the MLF one. This results in a misidentification of the mode in the SRF scenario (as the identification is based on the MLF modes) where the leading mode is found to be antisymmetric while a symmetric mode is expected in this short wavelength regime. Convergence issues with respect to the grid sampling or vacuum size might be the cause for this as the exponential in the integral of the SRF (Eq.~\ref{eq:srf}) can become quite important at large wavevector and large vacuum.

The analysis above shows the complexity of the interplay between different plasmonic modes. It also highlights the importance of the choice of the response function.

\section{Conclusion}\label{conclusion}

In this study, we have investigated the dielectric properties of jellium slabs, focusing on both single-slab and double-slab configurations. 
Through a combination of theoretical modeling and computational analysis, we have uncovered valuable insights into the behavior of surface modes and collective excitations in these systems. 
Our tool could be used to extend this analysis to many different systems.

Our analysis of single slabs revealed the impact of barrier height of the NSCF approach on the dispersion curves and loss functions. 
The right choice of this parameter can alleviate the need to use a self-consistent approach when the surface dielectric properties are considered in this single-slab case. 
The curves obtained from NSCF and SCF approaches for the symmetric and antisymmetric surface modes are very close to the ones of first-principles computations and even better in the long wavevelength regime where the size of the supercell is difficult to converge. 
We also highlight the failure of the macroscopic loss function to show the antisymmetric mode in symmetric systems in both NSCF and SCF approaches. 
Conversely, the surface response function presents all the modes.

Extending our investigation to two slabs introduced a new complexity, with the spacing between the slabs influencing the behavior of surface modes. We identified four distinct modes arising from interactions between symmetric slabs, shedding light on the interplay between surface-plasmon excitations in multi-layered structures. In asymmetric slabs, coupling between modes of different symmetries is observed under certain constraints such as having densities not too different (here this threshold is a variation of more than 0.02 $\si{\angstrom}^{-3}$ between the two slabs, but it depends on the SPP frequency of each material) and wavevectors small enough (this value depends on the difference of density).  

These findings offer valuable insights for the design and characterization of nanoscale devices and materials, where surface plasmonics plays a crucial role in determining optical properties and device performance. Moreover, our study opens up new avenues for future research, including the exploration of more complex excitations, the development of advanced modeling techniques, and the investigation of novel materials and geometries. Especially the search for coupling of plasmon modes in nanostructures of different materials is an interesting path.

In conclusion, our work contributes to the growing body of knowledge on surface plasmonics in jellium systems, offering valuable insights into their optical properties and paving the way for future advancements in nanophotonics and materials science.

%%%%%%%%%%%%%%%%%%%%%%%%%%
%\makeatletter
%\renewcommand \thesection{S\@arabic\c@section}
%\renewcommand\thetable{S\@arabic\c@table}
%\renewcommand \thefigure{S\@arabic\c@figure}
%\makeatother
%\section{Appendix}

\begin{acknowledgments}
The authors acknowledge funding by the Federation Wallonia-Brussels
through the ARC SURFASCOPE (UCLouvain-UNamur, Convention 19/24-102).
Computational resources have been provided by the supercomputing
facilities of the Université catholique de Louvain
(CISM/UCL) and the Consortium des Equipements de Calcul
Intensif en Fédération Wallonie Bruxelles (CECI) funded by
the FRS-FNRS under Grant No. 2.5020.11.
\end{acknowledgments}

%
%\renewcommand\thefigure{A\arabic{figure}}
%\setcounter{figure}{0}
%\renewcommand\thetable{A\arabic{table}}
%\setcounter{table}{0}   
%\section{} 

%\newpage
\bibliography{main} 

%apsrev4-2.bst 2019-01-14 (MD) hand-edited version of apsrev4-1.bst
%Control: key (0)
%Control: author (72) initials jnrlst
%Control: editor formatted (1) identically to author
%Control: production of article title (-1) disabled
%Control: page (0) single
%Control: year (1) truncated
%Control: production of eprint (0) enabled
\begin{thebibliography}{71}%
\makeatletter
\providecommand \@ifxundefined [1]{%
 \@ifx{#1\undefined}
}%
\providecommand \@ifnum [1]{%
 \ifnum #1\expandafter \@firstoftwo
 \else \expandafter \@secondoftwo
 \fi
}%
\providecommand \@ifx [1]{%
 \ifx #1\expandafter \@firstoftwo
 \else \expandafter \@secondoftwo
 \fi
}%
\providecommand \natexlab [1]{#1}%
\providecommand \enquote  [1]{``#1''}%
\providecommand \bibnamefont  [1]{#1}%
\providecommand \bibfnamefont [1]{#1}%
\providecommand \citenamefont [1]{#1}%
\providecommand \href@noop [0]{\@secondoftwo}%
\providecommand \href [0]{\begingroup \@sanitize@url \@href}%
\providecommand \@href[1]{\@@startlink{#1}\@@href}%
\providecommand \@@href[1]{\endgroup#1\@@endlink}%
\providecommand \@sanitize@url [0]{\catcode `\\12\catcode `\$12\catcode `\&12\catcode `\#12\catcode `\^12\catcode `\_12\catcode `\%12\relax}%
\providecommand \@@startlink[1]{}%
\providecommand \@@endlink[0]{}%
\providecommand \url  [0]{\begingroup\@sanitize@url \@url }%
\providecommand \@url [1]{\endgroup\@href {#1}{\urlprefix }}%
\providecommand \urlprefix  [0]{URL }%
\providecommand \Eprint [0]{\href }%
\providecommand \doibase [0]{https://doi.org/}%
\providecommand \selectlanguage [0]{\@gobble}%
\providecommand \bibinfo  [0]{\@secondoftwo}%
\providecommand \bibfield  [0]{\@secondoftwo}%
\providecommand \translation [1]{[#1]}%
\providecommand \BibitemOpen [0]{}%
\providecommand \bibitemStop [0]{}%
\providecommand \bibitemNoStop [0]{.\EOS\space}%
\providecommand \EOS [0]{\spacefactor3000\relax}%
\providecommand \BibitemShut  [1]{\csname bibitem#1\endcsname}%
\let\auto@bib@innerbib\@empty
%</preamble>
\bibitem [{\citenamefont {Murray}\ and\ \citenamefont {Barnes}(2007)}]{murray2007plasmonic}%
  \BibitemOpen
  \bibfield  {author} {\bibinfo {author} {\bibfnamefont {W.~A.}\ \bibnamefont {Murray}}\ and\ \bibinfo {author} {\bibfnamefont {W.~L.}\ \bibnamefont {Barnes}},\ }\href@noop {} {\bibfield  {journal} {\bibinfo  {journal} {Advanced materials}\ }\textbf {\bibinfo {volume} {19}},\ \bibinfo {pages} {3771} (\bibinfo {year} {2007})}\BibitemShut {NoStop}%
\bibitem [{\citenamefont {Kravets}\ \emph {et~al.}(2018)\citenamefont {Kravets}, \citenamefont {Kabashin}, \citenamefont {Barnes},\ and\ \citenamefont {Grigorenko}}]{kravets2018plasmonic}%
  \BibitemOpen
  \bibfield  {author} {\bibinfo {author} {\bibfnamefont {V.~G.}\ \bibnamefont {Kravets}}, \bibinfo {author} {\bibfnamefont {A.~V.}\ \bibnamefont {Kabashin}}, \bibinfo {author} {\bibfnamefont {W.~L.}\ \bibnamefont {Barnes}},\ and\ \bibinfo {author} {\bibfnamefont {A.~N.}\ \bibnamefont {Grigorenko}},\ }\href@noop {} {\bibfield  {journal} {\bibinfo  {journal} {Chemical reviews}\ }\textbf {\bibinfo {volume} {118}},\ \bibinfo {pages} {5912} (\bibinfo {year} {2018})}\BibitemShut {NoStop}%
\bibitem [{\citenamefont {Hill}(2015)}]{hill2015plasmonic}%
  \BibitemOpen
  \bibfield  {author} {\bibinfo {author} {\bibfnamefont {R.~T.}\ \bibnamefont {Hill}},\ }\href@noop {} {\bibfield  {journal} {\bibinfo  {journal} {Wiley Interdisciplinary Reviews: Nanomedicine and Nanobiotechnology}\ }\textbf {\bibinfo {volume} {7}},\ \bibinfo {pages} {152} (\bibinfo {year} {2015})}\BibitemShut {NoStop}%
\bibitem [{\citenamefont {Ferry}\ \emph {et~al.}(2010)\citenamefont {Ferry}, \citenamefont {Munday},\ and\ \citenamefont {Atwater}}]{ferry2010design}%
  \BibitemOpen
  \bibfield  {author} {\bibinfo {author} {\bibfnamefont {V.~E.}\ \bibnamefont {Ferry}}, \bibinfo {author} {\bibfnamefont {J.~N.}\ \bibnamefont {Munday}},\ and\ \bibinfo {author} {\bibfnamefont {H.~A.}\ \bibnamefont {Atwater}},\ }\href@noop {} {\bibfield  {journal} {\bibinfo  {journal} {Advanced materials}\ }\textbf {\bibinfo {volume} {22}},\ \bibinfo {pages} {4794} (\bibinfo {year} {2010})}\BibitemShut {NoStop}%
\bibitem [{\citenamefont {Sharma}\ \emph {et~al.}(2012)\citenamefont {Sharma}, \citenamefont {Frontiera}, \citenamefont {Henry}, \citenamefont {Ringe},\ and\ \citenamefont {Van~Duyne}}]{sharma2012sers}%
  \BibitemOpen
  \bibfield  {author} {\bibinfo {author} {\bibfnamefont {B.}~\bibnamefont {Sharma}}, \bibinfo {author} {\bibfnamefont {R.~R.}\ \bibnamefont {Frontiera}}, \bibinfo {author} {\bibfnamefont {A.-I.}\ \bibnamefont {Henry}}, \bibinfo {author} {\bibfnamefont {E.}~\bibnamefont {Ringe}},\ and\ \bibinfo {author} {\bibfnamefont {R.~P.}\ \bibnamefont {Van~Duyne}},\ }\href@noop {} {\bibfield  {journal} {\bibinfo  {journal} {Materials today}\ }\textbf {\bibinfo {volume} {15}},\ \bibinfo {pages} {16} (\bibinfo {year} {2012})}\BibitemShut {NoStop}%
\bibitem [{\citenamefont {Willets}\ and\ \citenamefont {Van~Duyne}(2007)}]{willets2007localized}%
  \BibitemOpen
  \bibfield  {author} {\bibinfo {author} {\bibfnamefont {K.~A.}\ \bibnamefont {Willets}}\ and\ \bibinfo {author} {\bibfnamefont {R.~P.}\ \bibnamefont {Van~Duyne}},\ }\href@noop {} {\bibfield  {journal} {\bibinfo  {journal} {Annual review of physical chemistry}\ }\textbf {\bibinfo {volume} {58}},\ \bibinfo {pages} {267} (\bibinfo {year} {2007})}\BibitemShut {NoStop}%
\bibitem [{\citenamefont {Stiles}\ \emph {et~al.}(2008)\citenamefont {Stiles}, \citenamefont {Dieringer}, \citenamefont {Shah},\ and\ \citenamefont {Van~Duyne}}]{stiles2008surface}%
  \BibitemOpen
  \bibfield  {author} {\bibinfo {author} {\bibfnamefont {P.~L.}\ \bibnamefont {Stiles}}, \bibinfo {author} {\bibfnamefont {J.~A.}\ \bibnamefont {Dieringer}}, \bibinfo {author} {\bibfnamefont {N.~C.}\ \bibnamefont {Shah}},\ and\ \bibinfo {author} {\bibfnamefont {R.~P.}\ \bibnamefont {Van~Duyne}},\ }\href@noop {} {\bibfield  {journal} {\bibinfo  {journal} {Annu. Rev. Anal. Chem.}\ }\textbf {\bibinfo {volume} {1}},\ \bibinfo {pages} {601} (\bibinfo {year} {2008})}\BibitemShut {NoStop}%
\bibitem [{\citenamefont {Langer}\ \emph {et~al.}(2020)\citenamefont {Langer}, \citenamefont {Jimenez~de Aberasturi}, \citenamefont {Aizpurua}, \citenamefont {Alvarez-Puebla}, \citenamefont {Augui{\'e}}, \citenamefont {Baumberg}, \citenamefont {Bazan}, \citenamefont {Bell}, \citenamefont {Boisen}, \citenamefont {Brolo}, \citenamefont {Choo}, \citenamefont {Cialla-May}, \citenamefont {Deckert}, \citenamefont {Fabris}, \citenamefont {Faulds}, \citenamefont {Garc{\'\i}a~de Abajo}, \citenamefont {Goodacre}, \citenamefont {Graham}, \citenamefont {Haes}, \citenamefont {Haynes}, \citenamefont {Huck}, \citenamefont {Itoh}, \citenamefont {K{\"a}ll}, \citenamefont {Kneipp}, \citenamefont {Kotov}, \citenamefont {Kuang}, \citenamefont {Le~Ru}, \citenamefont {Lee}, \citenamefont {Li}, \citenamefont {Ling}, \citenamefont {Maier}, \citenamefont {Mayerh{\"o}fer}, \citenamefont {Moskovits}, \citenamefont {Murakoshi}, \citenamefont {Nam}, \citenamefont {Nie}, \citenamefont {Ozaki}, \citenamefont {Pastoriza-Santos},
  \citenamefont {Perez-Juste}, \citenamefont {Popp}, \citenamefont {Pucci}, \citenamefont {Reich}, \citenamefont {Ren}, \citenamefont {Schatz}, \citenamefont {Shegai}, \citenamefont {Schl{\"u}cker}, \citenamefont {Tay}, \citenamefont {Thomas}, \citenamefont {Tian}, \citenamefont {Van~Duyne}, \citenamefont {Vo-Dinh}, \citenamefont {Wang}, \citenamefont {Willets}, \citenamefont {Xu}, \citenamefont {Xu}, \citenamefont {Xu}, \citenamefont {Yamamoto}, \citenamefont {Zhao},\ and\ \citenamefont {Liz-Marz{\'a}n}}]{Langer2019}%
  \BibitemOpen
  \bibfield  {author} {\bibinfo {author} {\bibfnamefont {J.}~\bibnamefont {Langer}}, \bibinfo {author} {\bibfnamefont {D.}~\bibnamefont {Jimenez~de Aberasturi}}, \bibinfo {author} {\bibfnamefont {J.}~\bibnamefont {Aizpurua}}, \bibinfo {author} {\bibfnamefont {R.~A.}\ \bibnamefont {Alvarez-Puebla}}, \bibinfo {author} {\bibfnamefont {B.}~\bibnamefont {Augui{\'e}}}, \bibinfo {author} {\bibfnamefont {J.~J.}\ \bibnamefont {Baumberg}}, \bibinfo {author} {\bibfnamefont {G.~C.}\ \bibnamefont {Bazan}}, \bibinfo {author} {\bibfnamefont {S.~E.~J.}\ \bibnamefont {Bell}}, \bibinfo {author} {\bibfnamefont {A.}~\bibnamefont {Boisen}}, \bibinfo {author} {\bibfnamefont {A.~G.}\ \bibnamefont {Brolo}}, \bibinfo {author} {\bibfnamefont {J.}~\bibnamefont {Choo}}, \bibinfo {author} {\bibfnamefont {D.}~\bibnamefont {Cialla-May}}, \bibinfo {author} {\bibfnamefont {V.}~\bibnamefont {Deckert}}, \bibinfo {author} {\bibfnamefont {L.}~\bibnamefont {Fabris}}, \bibinfo {author} {\bibfnamefont {K.}~\bibnamefont {Faulds}}, \bibinfo {author}
  {\bibfnamefont {F.~J.}\ \bibnamefont {Garc{\'\i}a~de Abajo}}, \bibinfo {author} {\bibfnamefont {R.}~\bibnamefont {Goodacre}}, \bibinfo {author} {\bibfnamefont {D.}~\bibnamefont {Graham}}, \bibinfo {author} {\bibfnamefont {A.~J.}\ \bibnamefont {Haes}}, \bibinfo {author} {\bibfnamefont {C.~L.}\ \bibnamefont {Haynes}}, \bibinfo {author} {\bibfnamefont {C.}~\bibnamefont {Huck}}, \bibinfo {author} {\bibfnamefont {T.}~\bibnamefont {Itoh}}, \bibinfo {author} {\bibfnamefont {M.}~\bibnamefont {K{\"a}ll}}, \bibinfo {author} {\bibfnamefont {J.}~\bibnamefont {Kneipp}}, \bibinfo {author} {\bibfnamefont {N.~A.}\ \bibnamefont {Kotov}}, \bibinfo {author} {\bibfnamefont {H.}~\bibnamefont {Kuang}}, \bibinfo {author} {\bibfnamefont {E.~C.}\ \bibnamefont {Le~Ru}}, \bibinfo {author} {\bibfnamefont {H.~K.}\ \bibnamefont {Lee}}, \bibinfo {author} {\bibfnamefont {J.-F.}\ \bibnamefont {Li}}, \bibinfo {author} {\bibfnamefont {X.~Y.}\ \bibnamefont {Ling}}, \bibinfo {author} {\bibfnamefont {S.~A.}\ \bibnamefont {Maier}}, \bibinfo
  {author} {\bibfnamefont {T.}~\bibnamefont {Mayerh{\"o}fer}}, \bibinfo {author} {\bibfnamefont {M.}~\bibnamefont {Moskovits}}, \bibinfo {author} {\bibfnamefont {K.}~\bibnamefont {Murakoshi}}, \bibinfo {author} {\bibfnamefont {J.-M.}\ \bibnamefont {Nam}}, \bibinfo {author} {\bibfnamefont {S.}~\bibnamefont {Nie}}, \bibinfo {author} {\bibfnamefont {Y.}~\bibnamefont {Ozaki}}, \bibinfo {author} {\bibfnamefont {I.}~\bibnamefont {Pastoriza-Santos}}, \bibinfo {author} {\bibfnamefont {J.}~\bibnamefont {Perez-Juste}}, \bibinfo {author} {\bibfnamefont {J.}~\bibnamefont {Popp}}, \bibinfo {author} {\bibfnamefont {A.}~\bibnamefont {Pucci}}, \bibinfo {author} {\bibfnamefont {S.}~\bibnamefont {Reich}}, \bibinfo {author} {\bibfnamefont {B.}~\bibnamefont {Ren}}, \bibinfo {author} {\bibfnamefont {G.~C.}\ \bibnamefont {Schatz}}, \bibinfo {author} {\bibfnamefont {T.}~\bibnamefont {Shegai}}, \bibinfo {author} {\bibfnamefont {S.}~\bibnamefont {Schl{\"u}cker}}, \bibinfo {author} {\bibfnamefont {L.-L.}\ \bibnamefont {Tay}}, \bibinfo
  {author} {\bibfnamefont {K.~G.}\ \bibnamefont {Thomas}}, \bibinfo {author} {\bibfnamefont {Z.-Q.}\ \bibnamefont {Tian}}, \bibinfo {author} {\bibfnamefont {R.~P.}\ \bibnamefont {Van~Duyne}}, \bibinfo {author} {\bibfnamefont {T.}~\bibnamefont {Vo-Dinh}}, \bibinfo {author} {\bibfnamefont {Y.}~\bibnamefont {Wang}}, \bibinfo {author} {\bibfnamefont {K.~A.}\ \bibnamefont {Willets}}, \bibinfo {author} {\bibfnamefont {C.}~\bibnamefont {Xu}}, \bibinfo {author} {\bibfnamefont {H.}~\bibnamefont {Xu}}, \bibinfo {author} {\bibfnamefont {Y.}~\bibnamefont {Xu}}, \bibinfo {author} {\bibfnamefont {Y.~S.}\ \bibnamefont {Yamamoto}}, \bibinfo {author} {\bibfnamefont {B.}~\bibnamefont {Zhao}},\ and\ \bibinfo {author} {\bibfnamefont {L.~M.}\ \bibnamefont {Liz-Marz{\'a}n}},\ }\href {https://doi.org/10.1021/acsnano.9b04224} {\bibfield  {journal} {\bibinfo  {journal} {ACS Nano}\ }\textbf {\bibinfo {volume} {14}},\ \bibinfo {pages} {28} (\bibinfo {year} {2020})},\ \bibinfo {note} {pMID: 31478375},\ \Eprint
  {https://arxiv.org/abs/https://doi.org/10.1021/acsnano.9b04224} {https://doi.org/10.1021/acsnano.9b04224} \BibitemShut {NoStop}%
\bibitem [{\citenamefont {Neubrech}\ \emph {et~al.}(2017)\citenamefont {Neubrech}, \citenamefont {Huck}, \citenamefont {Weber}, \citenamefont {Pucci},\ and\ \citenamefont {Giessen}}]{Neubrech2017}%
  \BibitemOpen
  \bibfield  {author} {\bibinfo {author} {\bibfnamefont {F.}~\bibnamefont {Neubrech}}, \bibinfo {author} {\bibfnamefont {C.}~\bibnamefont {Huck}}, \bibinfo {author} {\bibfnamefont {K.}~\bibnamefont {Weber}}, \bibinfo {author} {\bibfnamefont {A.}~\bibnamefont {Pucci}},\ and\ \bibinfo {author} {\bibfnamefont {H.}~\bibnamefont {Giessen}},\ }\href {https://doi.org/10.1021/acs.chemrev.6b00743} {\bibfield  {journal} {\bibinfo  {journal} {Chemical Reviews}\ }\textbf {\bibinfo {volume} {117}},\ \bibinfo {pages} {5110} (\bibinfo {year} {2017})},\ \bibinfo {note} {pMID: 28358482},\ \Eprint {https://arxiv.org/abs/https://doi.org/10.1021/acs.chemrev.6b00743} {https://doi.org/10.1021/acs.chemrev.6b00743} \BibitemShut {NoStop}%
\bibitem [{\citenamefont {Colleu}\ \emph {et~al.}(2024)\citenamefont {Colleu}, \citenamefont {Fekete}, \citenamefont {Gonze}, \citenamefont {Cloots}, \citenamefont {Li{\'e}geois}, \citenamefont {Rignanese},\ and\ \citenamefont {Henrard}}]{colleu2024surface}%
  \BibitemOpen
  \bibfield  {author} {\bibinfo {author} {\bibfnamefont {T.}~\bibnamefont {Colleu}}, \bibinfo {author} {\bibfnamefont {A.}~\bibnamefont {Fekete}}, \bibinfo {author} {\bibfnamefont {X.}~\bibnamefont {Gonze}}, \bibinfo {author} {\bibfnamefont {A.}~\bibnamefont {Cloots}}, \bibinfo {author} {\bibfnamefont {V.}~\bibnamefont {Li{\'e}geois}}, \bibinfo {author} {\bibfnamefont {G.-M.}\ \bibnamefont {Rignanese}},\ and\ \bibinfo {author} {\bibfnamefont {L.}~\bibnamefont {Henrard}},\ }\href@noop {} {\bibfield  {journal} {\bibinfo  {journal} {Journal of Physics: Photonics}\ }\textbf {\bibinfo {volume} {6}},\ \bibinfo {pages} {025003} (\bibinfo {year} {2024})}\BibitemShut {NoStop}%
\bibitem [{\citenamefont {Barnes}\ \emph {et~al.}(2003)\citenamefont {Barnes}, \citenamefont {Dereux},\ and\ \citenamefont {Ebbesen}}]{barnes2003surface}%
  \BibitemOpen
  \bibfield  {author} {\bibinfo {author} {\bibfnamefont {W.~L.}\ \bibnamefont {Barnes}}, \bibinfo {author} {\bibfnamefont {A.}~\bibnamefont {Dereux}},\ and\ \bibinfo {author} {\bibfnamefont {T.~W.}\ \bibnamefont {Ebbesen}},\ }\href@noop {} {\bibfield  {journal} {\bibinfo  {journal} {nature}\ }\textbf {\bibinfo {volume} {424}},\ \bibinfo {pages} {824} (\bibinfo {year} {2003})}\BibitemShut {NoStop}%
\bibitem [{\citenamefont {Grigorenko}\ \emph {et~al.}(2012)\citenamefont {Grigorenko}, \citenamefont {Polini},\ and\ \citenamefont {Novoselov}}]{grigorenko2012graphene}%
  \BibitemOpen
  \bibfield  {author} {\bibinfo {author} {\bibfnamefont {A.~N.}\ \bibnamefont {Grigorenko}}, \bibinfo {author} {\bibfnamefont {M.}~\bibnamefont {Polini}},\ and\ \bibinfo {author} {\bibfnamefont {K.~S.}\ \bibnamefont {Novoselov}},\ }\href@noop {} {\bibfield  {journal} {\bibinfo  {journal} {Nature photonics}\ }\textbf {\bibinfo {volume} {6}},\ \bibinfo {pages} {749} (\bibinfo {year} {2012})}\BibitemShut {NoStop}%
\bibitem [{\citenamefont {Maier}\ \emph {et~al.}(2007)\citenamefont {Maier} \emph {et~al.}}]{maier2007plasmonics}%
  \BibitemOpen
  \bibfield  {author} {\bibinfo {author} {\bibfnamefont {S.~A.}\ \bibnamefont {Maier}} \emph {et~al.},\ }\href@noop {} {\emph {\bibinfo {title} {Plasmonics: fundamentals and applications}}},\ Vol.~\bibinfo {volume} {1}\ (\bibinfo  {publisher} {Springer},\ \bibinfo {year} {2007})\BibitemShut {NoStop}%
\bibitem [{\citenamefont {Ritchie}(1957)}]{ritchie1957plasma}%
  \BibitemOpen
  \bibfield  {author} {\bibinfo {author} {\bibfnamefont {R.~H.}\ \bibnamefont {Ritchie}},\ }\href@noop {} {\bibfield  {journal} {\bibinfo  {journal} {Physical review}\ }\textbf {\bibinfo {volume} {106}},\ \bibinfo {pages} {874} (\bibinfo {year} {1957})}\BibitemShut {NoStop}%
\bibitem [{\citenamefont {Eguiluz}\ \emph {et~al.}(1975)\citenamefont {Eguiluz}, \citenamefont {Ying},\ and\ \citenamefont {Quinn}}]{eguiluz1975influence}%
  \BibitemOpen
  \bibfield  {author} {\bibinfo {author} {\bibfnamefont {A.}~\bibnamefont {Eguiluz}}, \bibinfo {author} {\bibfnamefont {S.}~\bibnamefont {Ying}},\ and\ \bibinfo {author} {\bibfnamefont {J.}~\bibnamefont {Quinn}},\ }\href@noop {} {\bibfield  {journal} {\bibinfo  {journal} {Physical Review B}\ }\textbf {\bibinfo {volume} {11}},\ \bibinfo {pages} {2118} (\bibinfo {year} {1975})}\BibitemShut {NoStop}%
\bibitem [{\citenamefont {Eguiluz}(1980)}]{eguiluz1980electron}%
  \BibitemOpen
  \bibfield  {author} {\bibinfo {author} {\bibfnamefont {A.~G.}\ \bibnamefont {Eguiluz}},\ }\href@noop {} {\bibfield  {journal} {\bibinfo  {journal} {Solid State Communications}\ }\textbf {\bibinfo {volume} {33}},\ \bibinfo {pages} {21} (\bibinfo {year} {1980})}\BibitemShut {NoStop}%
\bibitem [{\citenamefont {Eguiluz}(1981)}]{eguiluz1981screening}%
  \BibitemOpen
  \bibfield  {author} {\bibinfo {author} {\bibfnamefont {A.~G.}\ \bibnamefont {Eguiluz}},\ }\href@noop {} {\bibfield  {journal} {\bibinfo  {journal} {Physical Review B}\ }\textbf {\bibinfo {volume} {23}},\ \bibinfo {pages} {1542} (\bibinfo {year} {1981})}\BibitemShut {NoStop}%
\bibitem [{\citenamefont {Andersen}\ \emph {et~al.}(2012)\citenamefont {Andersen}, \citenamefont {Jacobsen},\ and\ \citenamefont {Thygesen}}]{andersen2012spatially}%
  \BibitemOpen
  \bibfield  {author} {\bibinfo {author} {\bibfnamefont {K.}~\bibnamefont {Andersen}}, \bibinfo {author} {\bibfnamefont {K.~W.}\ \bibnamefont {Jacobsen}},\ and\ \bibinfo {author} {\bibfnamefont {K.~S.}\ \bibnamefont {Thygesen}},\ }\href@noop {} {\bibfield  {journal} {\bibinfo  {journal} {Physical Review B}\ }\textbf {\bibinfo {volume} {86}},\ \bibinfo {pages} {245129} (\bibinfo {year} {2012})}\BibitemShut {NoStop}%
\bibitem [{\citenamefont {Ferrell}(1957)}]{ferrell1957characteristic}%
  \BibitemOpen
  \bibfield  {author} {\bibinfo {author} {\bibfnamefont {R.~A.}\ \bibnamefont {Ferrell}},\ }\href@noop {} {\bibfield  {journal} {\bibinfo  {journal} {Physical Review}\ }\textbf {\bibinfo {volume} {107}},\ \bibinfo {pages} {450} (\bibinfo {year} {1957})}\BibitemShut {NoStop}%
\bibitem [{\citenamefont {Powell}\ and\ \citenamefont {Swan}(1959)}]{powell1959origin}%
  \BibitemOpen
  \bibfield  {author} {\bibinfo {author} {\bibfnamefont {C.}~\bibnamefont {Powell}}\ and\ \bibinfo {author} {\bibfnamefont {J.}~\bibnamefont {Swan}},\ }\href@noop {} {\bibfield  {journal} {\bibinfo  {journal} {Physical Review}\ }\textbf {\bibinfo {volume} {115}},\ \bibinfo {pages} {869} (\bibinfo {year} {1959})}\BibitemShut {NoStop}%
\bibitem [{\citenamefont {Economou}(1969)}]{economou1969surface}%
  \BibitemOpen
  \bibfield  {author} {\bibinfo {author} {\bibfnamefont {E.}~\bibnamefont {Economou}},\ }\href@noop {} {\bibfield  {journal} {\bibinfo  {journal} {Physical review}\ }\textbf {\bibinfo {volume} {182}},\ \bibinfo {pages} {539} (\bibinfo {year} {1969})}\BibitemShut {NoStop}%
\bibitem [{\citenamefont {Newns}(1970)}]{newns1970dielectric}%
  \BibitemOpen
  \bibfield  {author} {\bibinfo {author} {\bibfnamefont {D.}~\bibnamefont {Newns}},\ }\href@noop {} {\bibfield  {journal} {\bibinfo  {journal} {Physical Review B}\ }\textbf {\bibinfo {volume} {1}},\ \bibinfo {pages} {3304} (\bibinfo {year} {1970})}\BibitemShut {NoStop}%
\bibitem [{\citenamefont {Chen}\ and\ \citenamefont {Bolton}(1992)}]{chen1992retarded}%
  \BibitemOpen
  \bibfield  {author} {\bibinfo {author} {\bibfnamefont {M.}~\bibnamefont {Chen}}\ and\ \bibinfo {author} {\bibfnamefont {J.}~\bibnamefont {Bolton}},\ }\href@noop {} {\bibfield  {journal} {\bibinfo  {journal} {Superlattices and microstructures}\ }\textbf {\bibinfo {volume} {12}},\ \bibinfo {pages} {531} (\bibinfo {year} {1992})}\BibitemShut {NoStop}%
\bibitem [{\citenamefont {Dobson}(1992)}]{dobson1992electron}%
  \BibitemOpen
  \bibfield  {author} {\bibinfo {author} {\bibfnamefont {J.~F.}\ \bibnamefont {Dobson}},\ }\href@noop {} {\bibfield  {journal} {\bibinfo  {journal} {Physical Review B}\ }\textbf {\bibinfo {volume} {46}},\ \bibinfo {pages} {10163} (\bibinfo {year} {1992})}\BibitemShut {NoStop}%
\bibitem [{\citenamefont {Pitarke}\ \emph {et~al.}(2006)\citenamefont {Pitarke}, \citenamefont {Silkin}, \citenamefont {Chulkov},\ and\ \citenamefont {Echenique}}]{Pitarke:2006aa}%
  \BibitemOpen
  \bibfield  {author} {\bibinfo {author} {\bibfnamefont {J.~M.}\ \bibnamefont {Pitarke}}, \bibinfo {author} {\bibfnamefont {V.~M.}\ \bibnamefont {Silkin}}, \bibinfo {author} {\bibfnamefont {E.~V.}\ \bibnamefont {Chulkov}},\ and\ \bibinfo {author} {\bibfnamefont {P.~M.}\ \bibnamefont {Echenique}},\ }\href {https://doi.org/10.1088/0034-4885/70/1/r01} {\bibfield  {journal} {\bibinfo  {journal} {Reports on Progress in Physics}\ }\textbf {\bibinfo {volume} {70}},\ \bibinfo {pages} {1} (\bibinfo {year} {2006})}\BibitemShut {NoStop}%
\bibitem [{\citenamefont {Koval}\ \emph {et~al.}(2016)\citenamefont {Koval}, \citenamefont {Marchesin}, \citenamefont {Foerster},\ and\ \citenamefont {S{\'a}nchez-Portal}}]{Koval:2016aa}%
  \BibitemOpen
  \bibfield  {author} {\bibinfo {author} {\bibfnamefont {P.}~\bibnamefont {Koval}}, \bibinfo {author} {\bibfnamefont {F.}~\bibnamefont {Marchesin}}, \bibinfo {author} {\bibfnamefont {D.}~\bibnamefont {Foerster}},\ and\ \bibinfo {author} {\bibfnamefont {D.}~\bibnamefont {S{\'a}nchez-Portal}},\ }\href {https://doi.org/10.1088/0953-8984/28/21/214001} {\bibfield  {journal} {\bibinfo  {journal} {Journal of Physics: Condensed Matter}\ }\textbf {\bibinfo {volume} {28}},\ \bibinfo {pages} {214001} (\bibinfo {year} {2016})}\BibitemShut {NoStop}%
\bibitem [{\citenamefont {Titantah}\ and\ \citenamefont {Karttunen}(2016)}]{Titantah:2016aa}%
  \BibitemOpen
  \bibfield  {author} {\bibinfo {author} {\bibfnamefont {J.~T.}\ \bibnamefont {Titantah}}\ and\ \bibinfo {author} {\bibfnamefont {M.}~\bibnamefont {Karttunen}},\ }\href {https://doi.org/10.1140/epjb/e2016-70065-y} {\bibfield  {journal} {\bibinfo  {journal} {The European Physical Journal B}\ }\textbf {\bibinfo {volume} {89}},\ \bibinfo {pages} {125} (\bibinfo {year} {2016})}\BibitemShut {NoStop}%
\bibitem [{\citenamefont {Kuisma}\ \emph {et~al.}(2015)\citenamefont {Kuisma}, \citenamefont {Sakko}, \citenamefont {Rossi}, \citenamefont {Larsen}, \citenamefont {Enkovaara}, \citenamefont {Lehtovaara},\ and\ \citenamefont {Rantala}}]{Kuisma:2015aa}%
  \BibitemOpen
  \bibfield  {author} {\bibinfo {author} {\bibfnamefont {M.}~\bibnamefont {Kuisma}}, \bibinfo {author} {\bibfnamefont {A.}~\bibnamefont {Sakko}}, \bibinfo {author} {\bibfnamefont {T.~P.}\ \bibnamefont {Rossi}}, \bibinfo {author} {\bibfnamefont {A.~H.}\ \bibnamefont {Larsen}}, \bibinfo {author} {\bibfnamefont {J.}~\bibnamefont {Enkovaara}}, \bibinfo {author} {\bibfnamefont {L.}~\bibnamefont {Lehtovaara}},\ and\ \bibinfo {author} {\bibfnamefont {T.~T.}\ \bibnamefont {Rantala}},\ }\href {https://doi.org/10.1103/PhysRevB.91.115431} {\bibfield  {journal} {\bibinfo  {journal} {Physical Review B}\ }\textbf {\bibinfo {volume} {91}},\ \bibinfo {pages} {115431} (\bibinfo {year} {2015})}\BibitemShut {NoStop}%
\bibitem [{\citenamefont {Manninen}(1986)}]{manninen1986structures}%
  \BibitemOpen
  \bibfield  {author} {\bibinfo {author} {\bibfnamefont {M.}~\bibnamefont {Manninen}},\ }\href@noop {} {\bibfield  {journal} {\bibinfo  {journal} {Physical Review B}\ }\textbf {\bibinfo {volume} {34}},\ \bibinfo {pages} {6886} (\bibinfo {year} {1986})}\BibitemShut {NoStop}%
\bibitem [{\citenamefont {Lang}\ and\ \citenamefont {Kohn}(1970)}]{Lang:1970aa}%
  \BibitemOpen
  \bibfield  {author} {\bibinfo {author} {\bibfnamefont {N.~D.}\ \bibnamefont {Lang}}\ and\ \bibinfo {author} {\bibfnamefont {W.}~\bibnamefont {Kohn}},\ }\href {https://doi.org/10.1103/PhysRevB.1.4555} {\bibfield  {journal} {\bibinfo  {journal} {Physical Review B}\ }\textbf {\bibinfo {volume} {1}},\ \bibinfo {pages} {4555} (\bibinfo {year} {1970})}\BibitemShut {NoStop}%
\bibitem [{\citenamefont {Lang}\ and\ \citenamefont {Kohn}(1971)}]{Lang:1971aa}%
  \BibitemOpen
  \bibfield  {author} {\bibinfo {author} {\bibfnamefont {N.~D.}\ \bibnamefont {Lang}}\ and\ \bibinfo {author} {\bibfnamefont {W.}~\bibnamefont {Kohn}},\ }\href {https://doi.org/10.1103/PhysRevB.3.1215} {\bibfield  {journal} {\bibinfo  {journal} {Physical Review B}\ }\textbf {\bibinfo {volume} {3}},\ \bibinfo {pages} {1215} (\bibinfo {year} {1971})}\BibitemShut {NoStop}%
\bibitem [{\citenamefont {Asmar}\ and\ \citenamefont {Gwinn}(1992)}]{asmar1992jellium}%
  \BibitemOpen
  \bibfield  {author} {\bibinfo {author} {\bibfnamefont {N.}~\bibnamefont {Asmar}}\ and\ \bibinfo {author} {\bibfnamefont {E.}~\bibnamefont {Gwinn}},\ }\href@noop {} {\bibfield  {journal} {\bibinfo  {journal} {Physical Review B}\ }\textbf {\bibinfo {volume} {46}},\ \bibinfo {pages} {4752} (\bibinfo {year} {1992})}\BibitemShut {NoStop}%
\bibitem [{\citenamefont {Ekardt}(1984)}]{ekardt1984dynamical}%
  \BibitemOpen
  \bibfield  {author} {\bibinfo {author} {\bibfnamefont {W.}~\bibnamefont {Ekardt}},\ }\href@noop {} {\bibfield  {journal} {\bibinfo  {journal} {Physical review letters}\ }\textbf {\bibinfo {volume} {52}},\ \bibinfo {pages} {1925} (\bibinfo {year} {1984})}\BibitemShut {NoStop}%
\bibitem [{\citenamefont {Pitarke}\ and\ \citenamefont {Eguiluz}(2001)}]{pitarke2001jellium}%
  \BibitemOpen
  \bibfield  {author} {\bibinfo {author} {\bibfnamefont {J.}~\bibnamefont {Pitarke}}\ and\ \bibinfo {author} {\bibfnamefont {A.}~\bibnamefont {Eguiluz}},\ }\href@noop {} {\bibfield  {journal} {\bibinfo  {journal} {Physical Review B}\ }\textbf {\bibinfo {volume} {63}},\ \bibinfo {pages} {045116} (\bibinfo {year} {2001})}\BibitemShut {NoStop}%
\bibitem [{\citenamefont {Liebsch}(1997)}]{liebsch1997electronic}%
  \BibitemOpen
  \bibfield  {author} {\bibinfo {author} {\bibfnamefont {A.}~\bibnamefont {Liebsch}},\ }\href {https://books.google.be/books?id=mVsZnNHofOQC} {\emph {\bibinfo {title} {Electronic Excitations at Metal Surfaces}}},\ Language of science\ (\bibinfo  {publisher} {Springer},\ \bibinfo {year} {1997})\BibitemShut {NoStop}%
\bibitem [{\citenamefont {Dobson}\ and\ \citenamefont {Wang}(2004)}]{dobson2004testing}%
  \BibitemOpen
  \bibfield  {author} {\bibinfo {author} {\bibfnamefont {J.~F.}\ \bibnamefont {Dobson}}\ and\ \bibinfo {author} {\bibfnamefont {J.}~\bibnamefont {Wang}},\ }\href@noop {} {\bibfield  {journal} {\bibinfo  {journal} {Physical Review B}\ }\textbf {\bibinfo {volume} {69}},\ \bibinfo {pages} {235104} (\bibinfo {year} {2004})}\BibitemShut {NoStop}%
\bibitem [{\citenamefont {Echarri}\ \emph {et~al.}(2020)\citenamefont {Echarri}, \citenamefont {Skj{\o}lstrup}, \citenamefont {Pedersen},\ and\ \citenamefont {De~Abajo}}]{echarri2020theory}%
  \BibitemOpen
  \bibfield  {author} {\bibinfo {author} {\bibfnamefont {A.~R.}\ \bibnamefont {Echarri}}, \bibinfo {author} {\bibfnamefont {E.~J.~H.}\ \bibnamefont {Skj{\o}lstrup}}, \bibinfo {author} {\bibfnamefont {T.~G.}\ \bibnamefont {Pedersen}},\ and\ \bibinfo {author} {\bibfnamefont {F.~J.~G.}\ \bibnamefont {De~Abajo}},\ }\href@noop {} {\bibfield  {journal} {\bibinfo  {journal} {Physical Review Research}\ }\textbf {\bibinfo {volume} {2}},\ \bibinfo {pages} {023096} (\bibinfo {year} {2020})}\BibitemShut {NoStop}%
\bibitem [{\citenamefont {Rocca}\ \emph {et~al.}(1995)\citenamefont {Rocca}, \citenamefont {Yibing}, \citenamefont {de~Mongeot},\ and\ \citenamefont {Valbusa}}]{rocca1995surface}%
  \BibitemOpen
  \bibfield  {author} {\bibinfo {author} {\bibfnamefont {M.}~\bibnamefont {Rocca}}, \bibinfo {author} {\bibfnamefont {L.}~\bibnamefont {Yibing}}, \bibinfo {author} {\bibfnamefont {F.~B.}\ \bibnamefont {de~Mongeot}},\ and\ \bibinfo {author} {\bibfnamefont {U.}~\bibnamefont {Valbusa}},\ }\href@noop {} {\bibfield  {journal} {\bibinfo  {journal} {Physical Review B}\ }\textbf {\bibinfo {volume} {52}},\ \bibinfo {pages} {14947} (\bibinfo {year} {1995})}\BibitemShut {NoStop}%
\bibitem [{\citenamefont {Ritchie}\ and\ \citenamefont {Marusak}(1966)}]{ritchie1966surface}%
  \BibitemOpen
  \bibfield  {author} {\bibinfo {author} {\bibfnamefont {R.}~\bibnamefont {Ritchie}}\ and\ \bibinfo {author} {\bibfnamefont {A.}~\bibnamefont {Marusak}},\ }\href@noop {} {\bibfield  {journal} {\bibinfo  {journal} {Surface Science}\ }\textbf {\bibinfo {volume} {4}},\ \bibinfo {pages} {234} (\bibinfo {year} {1966})}\BibitemShut {NoStop}%
\bibitem [{\citenamefont {Vincent}\ and\ \citenamefont {Silcox}(1973)}]{vincent1973dispersion}%
  \BibitemOpen
  \bibfield  {author} {\bibinfo {author} {\bibfnamefont {R.}~\bibnamefont {Vincent}}\ and\ \bibinfo {author} {\bibfnamefont {J.}~\bibnamefont {Silcox}},\ }\href@noop {} {\bibfield  {journal} {\bibinfo  {journal} {Physical Review Letters}\ }\textbf {\bibinfo {volume} {31}},\ \bibinfo {pages} {1487} (\bibinfo {year} {1973})}\BibitemShut {NoStop}%
\bibitem [{\citenamefont {Tsuei}\ \emph {et~al.}(1991)\citenamefont {Tsuei}, \citenamefont {Plummer}, \citenamefont {Liebsch}, \citenamefont {Pehlke}, \citenamefont {Kempa},\ and\ \citenamefont {Bakshi}}]{tsuei1991normal}%
  \BibitemOpen
  \bibfield  {author} {\bibinfo {author} {\bibfnamefont {K.-D.}\ \bibnamefont {Tsuei}}, \bibinfo {author} {\bibfnamefont {E.}~\bibnamefont {Plummer}}, \bibinfo {author} {\bibfnamefont {A.}~\bibnamefont {Liebsch}}, \bibinfo {author} {\bibfnamefont {E.}~\bibnamefont {Pehlke}}, \bibinfo {author} {\bibfnamefont {K.}~\bibnamefont {Kempa}},\ and\ \bibinfo {author} {\bibfnamefont {P.}~\bibnamefont {Bakshi}},\ }\href@noop {} {\bibfield  {journal} {\bibinfo  {journal} {Surface science}\ }\textbf {\bibinfo {volume} {247}},\ \bibinfo {pages} {302} (\bibinfo {year} {1991})}\BibitemShut {NoStop}%
\bibitem [{\citenamefont {Penzar}\ and\ \citenamefont {Sunjic}(1984)}]{penzar1984surface}%
  \BibitemOpen
  \bibfield  {author} {\bibinfo {author} {\bibfnamefont {Z.}~\bibnamefont {Penzar}}\ and\ \bibinfo {author} {\bibfnamefont {M.}~\bibnamefont {Sunjic}},\ }\href@noop {} {\bibfield  {journal} {\bibinfo  {journal} {Physica Scripta}\ }\textbf {\bibinfo {volume} {30}},\ \bibinfo {pages} {431} (\bibinfo {year} {1984})}\BibitemShut {NoStop}%
\bibitem [{\citenamefont {Persson}\ and\ \citenamefont {Zaremba}(1985)}]{persson1985electron}%
  \BibitemOpen
  \bibfield  {author} {\bibinfo {author} {\bibfnamefont {B.}~\bibnamefont {Persson}}\ and\ \bibinfo {author} {\bibfnamefont {E.}~\bibnamefont {Zaremba}},\ }\href@noop {} {\bibfield  {journal} {\bibinfo  {journal} {Physical Review B}\ }\textbf {\bibinfo {volume} {31}},\ \bibinfo {pages} {1863} (\bibinfo {year} {1985})}\BibitemShut {NoStop}%
\bibitem [{\citenamefont {Garcia-Lekue}\ and\ \citenamefont {Pitarke}(2001)}]{garcia2001energy}%
  \BibitemOpen
  \bibfield  {author} {\bibinfo {author} {\bibfnamefont {A.}~\bibnamefont {Garcia-Lekue}}\ and\ \bibinfo {author} {\bibfnamefont {J.}~\bibnamefont {Pitarke}},\ }\href@noop {} {\bibfield  {journal} {\bibinfo  {journal} {Physical Review B}\ }\textbf {\bibinfo {volume} {64}},\ \bibinfo {pages} {035423} (\bibinfo {year} {2001})}\BibitemShut {NoStop}%
\bibitem [{\citenamefont {Schulte}(1976)}]{schulte1976theory}%
  \BibitemOpen
  \bibfield  {author} {\bibinfo {author} {\bibfnamefont {F.}~\bibnamefont {Schulte}},\ }\href@noop {} {\bibfield  {journal} {\bibinfo  {journal} {Surface Science}\ }\textbf {\bibinfo {volume} {55}},\ \bibinfo {pages} {427} (\bibinfo {year} {1976})}\BibitemShut {NoStop}%
\bibitem [{\citenamefont {Pitarke}\ and\ \citenamefont {Eguiluz}(1998)}]{pitarke1998surface}%
  \BibitemOpen
  \bibfield  {author} {\bibinfo {author} {\bibfnamefont {J.}~\bibnamefont {Pitarke}}\ and\ \bibinfo {author} {\bibfnamefont {A.}~\bibnamefont {Eguiluz}},\ }\href@noop {} {\bibfield  {journal} {\bibinfo  {journal} {Physical Review B}\ }\textbf {\bibinfo {volume} {57}},\ \bibinfo {pages} {6329} (\bibinfo {year} {1998})}\BibitemShut {NoStop}%
\bibitem [{\citenamefont {Silkin}\ \emph {et~al.}(2004)\citenamefont {Silkin}, \citenamefont {Chulkov},\ and\ \citenamefont {Echenique}}]{silkin2004band}%
  \BibitemOpen
  \bibfield  {author} {\bibinfo {author} {\bibfnamefont {V.~M.}\ \bibnamefont {Silkin}}, \bibinfo {author} {\bibfnamefont {E.~V.}\ \bibnamefont {Chulkov}},\ and\ \bibinfo {author} {\bibfnamefont {P.~M.}\ \bibnamefont {Echenique}},\ }\href@noop {} {\bibfield  {journal} {\bibinfo  {journal} {Physical review letters}\ }\textbf {\bibinfo {volume} {93}},\ \bibinfo {pages} {176801} (\bibinfo {year} {2004})}\BibitemShut {NoStop}%
\bibitem [{\citenamefont {Silkin}\ \emph {et~al.}(2011)\citenamefont {Silkin}, \citenamefont {Nagao}, \citenamefont {Despoja}, \citenamefont {Echeverry}, \citenamefont {Eremeev}, \citenamefont {Chulkov},\ and\ \citenamefont {Echenique}}]{silkin2011low}%
  \BibitemOpen
  \bibfield  {author} {\bibinfo {author} {\bibfnamefont {V.~M.}\ \bibnamefont {Silkin}}, \bibinfo {author} {\bibfnamefont {T.}~\bibnamefont {Nagao}}, \bibinfo {author} {\bibfnamefont {V.}~\bibnamefont {Despoja}}, \bibinfo {author} {\bibfnamefont {J.}~\bibnamefont {Echeverry}}, \bibinfo {author} {\bibfnamefont {S.}~\bibnamefont {Eremeev}}, \bibinfo {author} {\bibfnamefont {E.~V.}\ \bibnamefont {Chulkov}},\ and\ \bibinfo {author} {\bibfnamefont {P.~M.}\ \bibnamefont {Echenique}},\ }\href@noop {} {\bibfield  {journal} {\bibinfo  {journal} {Physical Review B}\ }\textbf {\bibinfo {volume} {84}},\ \bibinfo {pages} {165416} (\bibinfo {year} {2011})}\BibitemShut {NoStop}%
\bibitem [{\citenamefont {Schaich}\ and\ \citenamefont {Dobson}(1994)}]{schaich1994excitation}%
  \BibitemOpen
  \bibfield  {author} {\bibinfo {author} {\bibfnamefont {W.}~\bibnamefont {Schaich}}\ and\ \bibinfo {author} {\bibfnamefont {J.~F.}\ \bibnamefont {Dobson}},\ }\href@noop {} {\bibfield  {journal} {\bibinfo  {journal} {Physical Review B}\ }\textbf {\bibinfo {volume} {49}},\ \bibinfo {pages} {14700} (\bibinfo {year} {1994})}\BibitemShut {NoStop}%
\bibitem [{\citenamefont {Giorgetti}\ \emph {et~al.}(2020)\citenamefont {Giorgetti}, \citenamefont {Iagupov},\ and\ \citenamefont {V{\'e}niard}}]{giorgetti2020electron}%
  \BibitemOpen
  \bibfield  {author} {\bibinfo {author} {\bibfnamefont {C.}~\bibnamefont {Giorgetti}}, \bibinfo {author} {\bibfnamefont {I.}~\bibnamefont {Iagupov}},\ and\ \bibinfo {author} {\bibfnamefont {V.}~\bibnamefont {V{\'e}niard}},\ }\href@noop {} {\bibfield  {journal} {\bibinfo  {journal} {Physical Review B}\ }\textbf {\bibinfo {volume} {101}},\ \bibinfo {pages} {035431} (\bibinfo {year} {2020})}\BibitemShut {NoStop}%
\bibitem [{\citenamefont {Despoja}\ \emph {et~al.}(2011)\citenamefont {Despoja}, \citenamefont {Echenique},\ and\ \citenamefont {{\v{S}}unji{\'c}}}]{despoja2011nonlocal}%
  \BibitemOpen
  \bibfield  {author} {\bibinfo {author} {\bibfnamefont {V.}~\bibnamefont {Despoja}}, \bibinfo {author} {\bibfnamefont {P.~M.}\ \bibnamefont {Echenique}},\ and\ \bibinfo {author} {\bibfnamefont {M.}~\bibnamefont {{\v{S}}unji{\'c}}},\ }\href@noop {} {\bibfield  {journal} {\bibinfo  {journal} {Physical Review B}\ }\textbf {\bibinfo {volume} {83}},\ \bibinfo {pages} {205424} (\bibinfo {year} {2011})}\BibitemShut {NoStop}%
\bibitem [{\citenamefont {Despoja}\ \emph {et~al.}(2006)\citenamefont {Despoja}, \citenamefont {Maru{\v{s}}i{\'c}},\ and\ \citenamefont {{\v{S}}unji{\'c}}}]{despoja2006excitation}%
  \BibitemOpen
  \bibfield  {author} {\bibinfo {author} {\bibfnamefont {V.}~\bibnamefont {Despoja}}, \bibinfo {author} {\bibfnamefont {L.}~\bibnamefont {Maru{\v{s}}i{\'c}}},\ and\ \bibinfo {author} {\bibfnamefont {M.}~\bibnamefont {{\v{S}}unji{\'c}}},\ }\href@noop {} {\bibfield  {journal} {\bibinfo  {journal} {Solid state communications}\ }\textbf {\bibinfo {volume} {140}},\ \bibinfo {pages} {270} (\bibinfo {year} {2006})}\BibitemShut {NoStop}%
\bibitem [{\citenamefont {Politano}\ \emph {et~al.}(2008)\citenamefont {Politano}, \citenamefont {Formoso},\ and\ \citenamefont {Chiarello}}]{politano2008dispersion}%
  \BibitemOpen
  \bibfield  {author} {\bibinfo {author} {\bibfnamefont {A.}~\bibnamefont {Politano}}, \bibinfo {author} {\bibfnamefont {V.}~\bibnamefont {Formoso}},\ and\ \bibinfo {author} {\bibfnamefont {G.}~\bibnamefont {Chiarello}},\ }\href@noop {} {\bibfield  {journal} {\bibinfo  {journal} {Plasmonics}\ }\textbf {\bibinfo {volume} {3}},\ \bibinfo {pages} {165} (\bibinfo {year} {2008})}\BibitemShut {NoStop}%
\bibitem [{\citenamefont {Giuliani}\ and\ \citenamefont {Vignale}(2008)}]{giuliani2008quantum}%
  \BibitemOpen
  \bibfield  {author} {\bibinfo {author} {\bibfnamefont {G.}~\bibnamefont {Giuliani}}\ and\ \bibinfo {author} {\bibfnamefont {G.}~\bibnamefont {Vignale}},\ }\href@noop {} {\emph {\bibinfo {title} {Quantum theory of the electron liquid}}}\ (\bibinfo  {publisher} {Cambridge university press},\ \bibinfo {year} {2008})\BibitemShut {NoStop}%
\bibitem [{\citenamefont {Despoja}\ \emph {et~al.}(2005)\citenamefont {Despoja}, \citenamefont {Maru{\v{s}}i{\'c}},\ and\ \citenamefont {{\v{S}}unji{\'c}}}]{Despoja:2005aa}%
  \BibitemOpen
  \bibfield  {author} {\bibinfo {author} {\bibfnamefont {V.}~\bibnamefont {Despoja}}, \bibinfo {author} {\bibfnamefont {L.}~\bibnamefont {Maru{\v{s}}i{\'c}}},\ and\ \bibinfo {author} {\bibfnamefont {M.}~\bibnamefont {{\v{S}}unji{\'c}}},\ }\href@noop {} {\bibfield  {journal} {\bibinfo  {journal} {Fizika A}\ }\textbf {\bibinfo {volume} {14}},\ \bibinfo {pages} {207} (\bibinfo {year} {2005})}\BibitemShut {NoStop}%
\bibitem [{\citenamefont {Gonze}(1996)}]{gonze1996towards}%
  \BibitemOpen
  \bibfield  {author} {\bibinfo {author} {\bibfnamefont {X.}~\bibnamefont {Gonze}},\ }\href@noop {} {\bibfield  {journal} {\bibinfo  {journal} {Physical Review B}\ }\textbf {\bibinfo {volume} {54}},\ \bibinfo {pages} {4383} (\bibinfo {year} {1996})}\BibitemShut {NoStop}%
\bibitem [{\citenamefont {Eguiluz}(1985)}]{eguiluz1985self}%
  \BibitemOpen
  \bibfield  {author} {\bibinfo {author} {\bibfnamefont {A.~G.}\ \bibnamefont {Eguiluz}},\ }\href@noop {} {\bibfield  {journal} {\bibinfo  {journal} {Physical Review B}\ }\textbf {\bibinfo {volume} {31}},\ \bibinfo {pages} {3303} (\bibinfo {year} {1985})}\BibitemShut {NoStop}%
\bibitem [{1()}]{1}%
  \BibitemOpen
  \href@noop {} {\bibinfo {title} {Thanks to the periodicity of the problem, the response functions usually expressed as functions of $\bs{r}$ and $\bs{r'}$ are here described by a vector $q_{||}$ invariant rotational in the plane $x-y$ and two spatial coordinates $z, z'$. to go from the former to the latter, a 2d fourier transform must be performed on the $x-y$ coordinates.}}\BibitemShut {Stop}%
\bibitem [{2()}]{2}%
  \BibitemOpen
  \href@noop {} {\bibinfo {title} {The sum over $k_{||}$ can be replaced by a factor $2a(e_f-\epsilon_l)/(2\pi)$.}}\BibitemShut {Stop}%
\bibitem [{3()}]{3}%
  \BibitemOpen
  \href@noop {} {\bibinfo {title} {The screened interaction being computed from eq.~(\ref{eq:mat_W}), the response can be analyzed over the whole space, not only for $z, z'$ pairs far from the surface as suggested by eq.~(\ref{eq:g_to_W}).}}\BibitemShut {Stop}%
\bibitem [{4()}]{4}%
  \BibitemOpen
  \href@noop {} {\bibinfo {title} {The simple fourier transform is defined for a quantity $\alpha(z, z',q_{||}, \omega)$ as $\alpha(q_{||}, g) = 1/l\int\exp[-i(q_{||}+g)z]\alpha(z, z',q_{||}, \omega))dz$ while the double fourier transform of a quantity $\beta(z, z',q_{||}, \omega)$ is defined as $\beta(q_{||}, g, g', \omega) = 1/l^2\int\int\exp[-i(q+g)z]\beta(z, z',q_{||}, \omega)\exp[i(q_{||}+g')z']dzdz'$.}}\BibitemShut {Stop}%
\bibitem [{\citenamefont {Maj{\'e}rus}\ \emph {et~al.}(2023)\citenamefont {Maj{\'e}rus}, \citenamefont {Guillaume}, \citenamefont {Kockaert},\ and\ \citenamefont {Henrard}}]{majerus2023anisotropy}%
  \BibitemOpen
  \bibfield  {author} {\bibinfo {author} {\bibfnamefont {B.}~\bibnamefont {Maj{\'e}rus}}, \bibinfo {author} {\bibfnamefont {E.}~\bibnamefont {Guillaume}}, \bibinfo {author} {\bibfnamefont {P.}~\bibnamefont {Kockaert}},\ and\ \bibinfo {author} {\bibfnamefont {L.}~\bibnamefont {Henrard}},\ }\href@noop {} {\bibfield  {journal} {\bibinfo  {journal} {Physical Review B}\ }\textbf {\bibinfo {volume} {108}},\ \bibinfo {pages} {245412} (\bibinfo {year} {2023})}\BibitemShut {NoStop}%
\bibitem [{5()}]{5}%
  \BibitemOpen
  \href@noop {} {\bibinfo {title} {The intensities can be compared in normalized spectra $l\ind{slab}(\omega)$ with $l\ind{slab}(\omega) = l\ind{M}(\omega)\frac{d\ind{vac}}{d\ind{cell}}$ with $d\ind{vac}$ and $d\ind{cell}$ the width of the vacuum and of the total cell (vacuum and jellium) respectively.}}\BibitemShut {Stop}%
\bibitem [{\citenamefont {Friedel}(1952)}]{friedel1952xiv}%
  \BibitemOpen
  \bibfield  {author} {\bibinfo {author} {\bibfnamefont {J.}~\bibnamefont {Friedel}},\ }\href@noop {} {\bibfield  {journal} {\bibinfo  {journal} {The London, Edinburgh, and Dublin Philosophical Magazine and Journal of Science}\ }\textbf {\bibinfo {volume} {43}},\ \bibinfo {pages} {153} (\bibinfo {year} {1952})}\BibitemShut {NoStop}%
\bibitem [{\citenamefont {Wigner}\ and\ \citenamefont {Bardeen}(1997)}]{wigner1997theory}%
  \BibitemOpen
  \bibfield  {author} {\bibinfo {author} {\bibfnamefont {E.}~\bibnamefont {Wigner}}\ and\ \bibinfo {author} {\bibfnamefont {J.}~\bibnamefont {Bardeen}},\ }\href@noop {} {\bibfield  {journal} {\bibinfo  {journal} {Part I: Physical Chemistry. Part II: Solid State Physics}\ ,\ \bibinfo {pages} {398}} (\bibinfo {year} {1997})}\BibitemShut {NoStop}%
\bibitem [{\citenamefont {Ashcroft}\ and\ \citenamefont {Mermin}(1976)}]{burchamashcroft}%
  \BibitemOpen
  \bibfield  {author} {\bibinfo {author} {\bibfnamefont {N.}~\bibnamefont {Ashcroft}}\ and\ \bibinfo {author} {\bibfnamefont {N.}~\bibnamefont {Mermin}},\ }\href@noop {} {\emph {\bibinfo {title} {Solid State Physics}}}\ (\bibinfo  {publisher} {Saunders, Philadelphia},\ \bibinfo {year} {1976})\BibitemShut {NoStop}%
\bibitem [{\citenamefont {Echarri}\ \emph {et~al.}(2021)\citenamefont {Echarri}, \citenamefont {Gon{\c{c}}alves}, \citenamefont {Tserkezis}, \citenamefont {de~Abajo}, \citenamefont {Mortensen},\ and\ \citenamefont {Cox}}]{echarri2021optical}%
  \BibitemOpen
  \bibfield  {author} {\bibinfo {author} {\bibfnamefont {A.~R.}\ \bibnamefont {Echarri}}, \bibinfo {author} {\bibfnamefont {P.}~\bibnamefont {Gon{\c{c}}alves}}, \bibinfo {author} {\bibfnamefont {C.}~\bibnamefont {Tserkezis}}, \bibinfo {author} {\bibfnamefont {F.~J.~G.}\ \bibnamefont {de~Abajo}}, \bibinfo {author} {\bibfnamefont {N.~A.}\ \bibnamefont {Mortensen}},\ and\ \bibinfo {author} {\bibfnamefont {J.~D.}\ \bibnamefont {Cox}},\ }\href@noop {} {\bibfield  {journal} {\bibinfo  {journal} {Optica}\ }\textbf {\bibinfo {volume} {8}},\ \bibinfo {pages} {710} (\bibinfo {year} {2021})}\BibitemShut {NoStop}%
\bibitem [{\citenamefont {Cirac{\`\i}}\ and\ \citenamefont {Della~Sala}(2016)}]{ciraci2016quantum}%
  \BibitemOpen
  \bibfield  {author} {\bibinfo {author} {\bibfnamefont {C.}~\bibnamefont {Cirac{\`\i}}}\ and\ \bibinfo {author} {\bibfnamefont {F.}~\bibnamefont {Della~Sala}},\ }\href@noop {} {\bibfield  {journal} {\bibinfo  {journal} {Physical Review B}\ }\textbf {\bibinfo {volume} {93}},\ \bibinfo {pages} {205405} (\bibinfo {year} {2016})}\BibitemShut {NoStop}%
\bibitem [{6()}]{6}%
  \BibitemOpen
  \href@noop {} {\bibinfo {title} {In the paper of echarri \textit{et al.}\cite{echarri2020theory}, their dispersion curves does not show this trend but we believe they did not inspect sufficiently short-wavelength for it to appear, as they limit their study to wavevector up to 0.1~\si{\angstrom}$^{-1}$.}}\BibitemShut {Stop}%
\bibitem [{\citenamefont {Richter}\ and\ \citenamefont {Geiger}(1981)}]{richter1981energy}%
  \BibitemOpen
  \bibfield  {author} {\bibinfo {author} {\bibfnamefont {H.}~\bibnamefont {Richter}}\ and\ \bibinfo {author} {\bibfnamefont {J.}~\bibnamefont {Geiger}},\ }\href@noop {} {\bibfield  {journal} {\bibinfo  {journal} {Zeitschrift f{\"u}r Physik B Condensed Matter}\ }\textbf {\bibinfo {volume} {42}},\ \bibinfo {pages} {39} (\bibinfo {year} {1981})}\BibitemShut {NoStop}%
\bibitem [{7()}]{7}%
  \BibitemOpen
  \href@noop {} {\bibinfo {title} {The symbol used for the hyperbolic tangent and cotangent function in the original paper are misleading: the 'tan' and 'cot' symbol should be replaced by 'tanh' and 'coth'.}}\BibitemShut {Stop}%
\end{thebibliography}%

\end{document}